\documentclass[traditabstract]{aa}
\pdfoutput=1
\usepackage{verbatim} 
\usepackage{amsmath}
\usepackage{amssymb}
\usepackage{graphicx}
\usepackage{booktabs}
\usepackage[authoryear]{natbib}
\usepackage[colorlinks=true,linkcolor=blue,urlcolor=blue,citecolor=blue]{hyperref}
\usepackage{xspace}

\makeatletter

\usepackage{txfonts}\usepackage{twoopt}
\usepackage{lscape}
\usepackage{longtable}
\usepackage{threeparttablex}
\bibpunct{(}{)}{;}{a}{}{,}
\usepackage{enumitem}

\newcommand{\msun}{{\rm M}_{\odot}}
\newcommand{\lsun}{{\rm L}_{\odot}}

\newcommand{\kms}{{\rm km\,s^{-1}}}
\newcommand{\myr}{{\rm Myr}}
\newcommand{\bonnsai}{\mbox{\textsc{Bonnsai}}\xspace}

\newcommand{\tdor}{\mbox{30~Dor}\xspace}
\newcommand{\flames}{\mbox{\uppercase{Flames}}\xspace}
\newcommand{\argus}{\mbox{\uppercase{Argus}}\xspace}
\newcommand{\gaia}{\mbox{\it Gaia}\xspace}
\newcommand{\tld}{\mbox{TLD\,1}\xspace}
\newcommand{\pulsar}{\mbox{PSR~J0537$-$6910}\xspace}
\newcommand{\eg}{e.g.\@\xspace}
\newcommand{\cf}{c.f.\@\xspace}
\newcommand{\ie}{i.e.\@\xspace}

\titlerunning{Massive star formation in the local 30~Doradus starburst}
\authorrunning{F.R.N.~Schneider~et~al.}
\makeatother
\begin{document}
\title{The VLT-FLAMES Tarantula Survey.\thanks{Based on observations collected at the European Southern Observatory under programme ID 182.D-0222.}}
\subtitle{XXIX. Massive star formation in the local 30~Doradus starburst} 

\author{F.R.N.~Schneider\inst{\ref{OXFORD}}\thanks{fabian.schneider@physics.ox.ac.uk} 
\and O.H. Ram{\'i}rez-Agudelo\inst{\ref{Edin}}
\and F. Tramper\inst{\ref{Madrid}}
\and J.M. Bestenlehner\inst{\ref{MPIA},\ref{Sheffield}}
\and N. Castro\inst{\ref{MICH}}
\and H. Sana\inst{\ref{Leuven}}
\and C.J. Evans\inst{\ref{Edin}}
\and C. Sab{\'i}n-Sanjuli{\'a}n\inst{\ref{Chile}}
\and S. Sim{\'o}n-D{\'i}az\inst{\ref{IAC},\ref{DA}}
\and N. Langer\inst{\ref{AIFA}}
\and L. Fossati\inst{\ref{Graz}}
\and G. Gr{\"a}fener\inst{\ref{AIFA}}
\and P.A. Crowther\inst{\ref{Sheffield}}
\and S.E. de Mink\inst{\ref{Amsterdam}}
\and A. de Koter\inst{\ref{Amsterdam},\ref{Leuven}}
\and M. Gieles\inst{\ref{Surrey}}
\and A. Herrero\inst{\ref{IAC},\ref{DA}}
\and R.G. Izzard\inst{\ref{Surrey},\ref{IOA}}
\and V. Kalari\inst{\ref{Chile2}}
\and R.S. Klessen\inst{\ref{ITA}}
\and D.J. Lennon\inst{\ref{Madrid}}
\and L. Mahy\inst{\ref{Leuven}}
\and J. Ma\'{i}z Apell\'{a}niz\inst{\ref{ESAC}}
\and N. Markova\inst{\ref{Bulgaria}}
\and J.Th. van Loon\inst{\ref{Keele}}
\and J.S. Vink\inst{\ref{Armagh}}
\and N.R. Walborn\inst{\ref{STScI}}\thanks{We regret to inform that Dr Nolan Walborn passed away earlier this year.}}
\institute{Department of Physics, University of Oxford, Denys Wilkinson Building, Keble Road, Oxford OX1 3RH, United Kingdom\label{OXFORD}
\and UK Astronomy Technology Centre, Royal Observatory Edinburgh, Blackford Hill, Edinburgh EH9 3HJ, United Kingdom\label{Edin}
\and European Space Astronomy Centre, Mission Operations Division, PO Box 78, 28691 Villanueva de la Ca\~nada, Madrid, Spain\label{Madrid}
\and Max-Planck-Institut f{\"u}r Astronomie, K{\"o}nigstuhl 17, 69117 Heidelberg, Germany\label{MPIA}
\and Department of Physics and Astronomy, Hicks Building, Hounsfield Road, University of Sheffield, Sheffield S3 7RH, United Kingdom\label{Sheffield}
\and Department of Astronomy, University of Michigan, 1085 S. University Avenue, Ann Arbor, MI 48109-1107, USA\label{MICH}
\and Institute of Astrophysics, KU Leuven, Celestijnenlaan 200D, 3001, Leuven, Belgium\label{Leuven} 
\and Departamento de F{\'i}sica y Astronom{\'i}a, Universidad de La Serena, Avda. Juan Cisternas $N^o$ 1200 Norte, La Serena, Chile\label{Chile}
\and Instituto de Astrof{\'i}sica de Canarias, E-38205 La Laguna, Tenerife, Spain\label{IAC}
\and Departamento de Astrof{\'i}sica, Universidad de La Laguna, E-38206 La Laguna, Tenerife, Spain\label{DA}
\and Argelander-Institut f{\"u}r Astronomie der Universit{\"a}t Bonn, Auf dem H{\"u}gel~71, 53121~Bonn, Germany\label{AIFA}
\and Austrian Academy of Sciences, Space Research Institute, Schmiedlstra{\ss}e 6, 8042 Graz, Austria\label{Graz}
\and Astronomical Institute Anton Pannekoek, Amsterdam University, Science Park 904, 1098 XH Amsterdam, The Netherlands\label{Amsterdam}
\and Department of Physics, Faculty of Engineering and Physical Sciences, University of Surrey, Guildford, GU2 7XH, United Kingdom\label{Surrey}
\and Institute of Astronomy, The Observatories, Madingley Road, Cambridge CB3 0HA, United Kingdom\label{IOA}
\and Departamento de Astronom{\'i}a, Universidad de Chile, Camino El Observatorio 1515, Las Condes, Santiago, Casilla 36-D, Chile\label{Chile2}
\and Institut f{\"u}r Theoretische Astrophysik, Zentrum f{\"u}r Astronomie der Universit{\"a}t Heidelberg, Albert-Ueberle-Str. 2, 69120 Heidelberg, Germany\label{ITA}
\and Centro de Astrobiolog{\'i}a, CSIC-INTA, ESAC campus, camino bajo del castillo s/n, E-28\,692 Villanueva de la Ca\~nada, Spain\label{ESAC}
\and Institute of Astronomy with National Astronomical Observatory, Bulgarian Academy of Sciences, PO Box 136, 4700 Smoljan, Bulgaria\label{Bulgaria}
\and Lennard-Jones Laboratories, Keele University, Staffordshire, ST5 5BG, United Kingdom\label{Keele}
\and Armagh Observatory, College Hill, Armagh, BT61 9DG, Northern Ireland, United Kingdom\label{Armagh}
\and Space Telescope Science Institute, 3700 San Martin Drive, Baltimore, MD 21218, USA\label{STScI}
}
\date{Received 15 May 2018 / Accepted 03 July 2018}
\abstract{The 30~Doradus (\tdor) nebula in the Large Magellanic Cloud (LMC) is the brightest HII region in the Local Group and a prototype starburst similar to those found in high redshift galaxies. It is thus a stepping stone to understand the complex formation processes of stars in starburst regions across the Universe. Here, we have studied the formation history of massive stars in \tdor using masses and ages derived for 452 mainly OB stars from the spectroscopic VLT-FLAMES Tarantula Survey (VFTS). We find that stars of all ages and masses are scattered throughout \tdor. This is remarkable because it implies that massive stars either moved large distances or formed independently over the whole field of view in relative isolation. We find that both channels contribute to the \tdor massive star population. Massive star formation rapidly accelerated about $8\,\myr$ ago, first forming stars in the field before giving birth to the stellar populations in NGC~2060 and NGC~2070. The R136 star cluster in NGC~2070 formed last and, since then, about $1\,\myr$ ago, star formation seems to be diminished with some continuing in the surroundings of R136. Massive stars within a projected distance of $8\,\mathrm{pc}$ of R136 are not coeval but show an age range of up to $6\,\myr$. Our mass distributions are well populated up to $200\,\msun$. The inferred IMF is shallower than a Salpeter-like IMF and appears to be the same across \tdor. By comparing our sample of stars to stellar models in the Hertzsprung--Russell diagram, we find evidence for missing physics in the models above $\log L/\lsun=6$ that is likely connected to enhanced wind mass loss for stars approaching the Eddington limit. Our work highlights the key information about the formation, evolution and final fates of massive stars encapsulated in the stellar content of \tdor, and sets a new benchmark for theories of massive star formation in giant molecular clouds.}
\maketitle
\section{Introduction\label{sec:introduction}}

The Tarantula nebula, also known as 30~Doradus (\tdor), was once thought to be a single star---the 30th brightest in the constellation of Doradus---until Nicolas Louis de Lacaille realised its nebular structure in 1751 (see for example notes on \tdor, also known as NGC~2070, in the Messier and NGC catalogues; \href{http://www.messier.obspm.fr/xtra/ngc/n2070.html}{http://www.messier.obspm.fr/xtra/ngc/n2070.html}). Today we know that \tdor is a highly complex nebula and the brightest HII region in the Local Group \citep{1984ApJ...287..116K} illuminated by the central massive star cluster R136 that contains some of the most massive stars known to date \citep[$200\text{--}300\,\msun$;][]{2010MNRAS.408..731C,2011A&A...530L..14B,2014A&A...565A..27H}. Several other extreme objects such as the fastest rotating star VFTS~102 \citep{2011ApJ...743L..22D}, the fastest rotating and most energetic young pulsar \pulsar \citep{2006ApJ...651..237C} and the very massive runaway VFTS~016 \citep{2010ApJ...715L..74E} reside within \tdor. \citet{2018Sci...359...69S} have further shown that \tdor formed an excess of massive stars (${\geq}\,30\,\msun$) compared to a Salpeter-like stellar initial mass function \citep[IMF;][]{1955ApJ...121..161S}.

At a distance of $50\,\mathrm{kpc}$ \citep{2013Natur.495...76P}, \tdor is a unique star-forming region that allows us to study massive star evolution, star formation and cluster evolution in great detail. It is a template for distant, unresolved starbursts and can be used to explore their role in galaxies and the overall cosmos. 

For example, the integrated spectrum of the central NGC~2070 region shows similar nebular emission characteristics to so-called Green Pea galaxies \citep{2009MNRAS.399.1191C}, and its size and star-formation rate are comparable to knots and clumps of intense star formation in high redshift galaxies \citep{2018arXiv180100855C}. Some Green Pea galaxies are Lyman continuum leakers that have escape fractions of ionising radiation of up to 50\% \citep[\eg][]{2016Natur.529..178I,2017ApJ...851L...9J,2018MNRAS.474.4514I} and their high-redshift counterparts have therefore been suggested to play an important role for the reionisation of the Universe. \citet{2016A&A...590A..36C} model the radiation field and gas density in \tdor and find a highly porous interstellar medium surrounding R136 that allows hard photons to reach large distances from the ionising cluster. \citet{2013A&A...558A.134D} estimate an escape fraction of ionising photons from \tdor of $6^{+55}_{-6}\%$. 

The central R136 star cluster has further been suggested to be a young counterpart of relatively low-mass globular clusters \citep[\eg][]{1988AJ.....95..720K,1993ASPC...48..588M,1994ApJ...433...65O,1995ApJ...448..179H,2009ApJ...707.1347A,2009AJ....137.3437B} and the star formation and stellar populations in this region may provide insights for our understanding of multiple main-sequences and abundance anomalies observed in globular clusters \citep[see \eg][]{2012A&ARv..20...50G,2017arXiv171201286B,2018MNRAS.473L..11R}. \tdor therefore offers the unique possibility to understand the star formation process of giant starbursts across the Universe.

The \tdor nebula is a huge and complex star-forming region that has produced stars over the last ${\approx}30\,\myr$ \citep[\eg][]{1997ApJS..112..457W,2000AJ....119..787G,2011ApJ...739...27D,2015ApJ...811...76C}. Its main constituents are the star clusters R136 in the central NGC~2070 region, Hodge~301, SL~639 and NGC~2060, which further hosts the \tld star cluster. Star formation in \tdor is currently further explored in detail within the Hubble Tarantula Treasury Project \citep{2013AJ....146...53S,2015ApJ...811...76C,2016ApJS..222...11S,2018MNRAS.tmp.1265K}, a panchromatic imaging survey of \tdor using the Hubble Space Telescope, and the heart of \tdor has been dissected with the Multi Unit Spectroscopic Explorer (MUSE) on the Very Large Telescope \citep[VLT;][]{2018arXiv180100855C,2018A&A...614A.147C}.

Within the VLT-FLAMES Tarantula Survey \citep[VFTS;][]{2011A&A...530A.108E}, multi-epoch optical spectra of over 900 massive stars brighter than $V{=}17\,\mathrm{mag}$ have been obtained that offer the unique possibility to study individual objects and big samples of massive stars in \tdor in great detail. Of the VFTS \flames targets, 
\begin{itemize}[topsep=3pt]
\item 342 are classified as O stars \citep{2014A&A...564A..40W}, 
\item 438 as B stars \citep{2015A&A...574A..13E}, 
\item six as O2-3.5 If{*}/WN5-7, called `slash' stars from here on \citep{2011MNRAS.416.1311C}, 
\item 17 as Wolf--Rayet (WR) stars (two earlier and 12 later than WN5 and three WCs; see for example \citealp{2013A&A...558A.134D}, for the whole WR star population in \tdor) and 
\item 92 as cool-type stars \citep[A-type and later;][]{2011A&A...530A.108E}.
\end{itemize}
For a significant number of these massive stars (${>}\,500$), detailed atmosphere models have been computed, making this an unprecedented sample of early-type stars in one of the largest, resolved starburst regions in the local Universe. In combination with stellar evolutionary models and sophisticated statistical methods, \citet{2018Sci...359...69S} used this dataset to determine fundamental stellar parameters such as initial mass and age for each star, and thereby derive the star formation history and initial mass function in \tdor. Here, we utilise the inferred ages and masses to further our understanding of the star formation process in \tdor and in similar starbursts across cosmic history. 

This paper is structured as follows. In Sect.~\ref{sec:methods}, we briefly explain how the atmospheric parameters and the ages and masses of individual stars are determined. We further show how we compute age and mass distributions from these parameters to shed light on the star formation process in \tdor. In Sect.~\ref{sec:results}, we describe our sample stars with respect to stellar models, the inferred age and mass distributions of massive stars in various sub-regions of \tdor, the spatial distribution of stellar ages and masses, and the derived IMFs. The emerging picture of the overall star formation process is discussed in Sect.~\ref{sec:overall-sf-process} and our main conclusions are summarised in Sect.~\ref{sec:conclusions}.

\section{Methods}\label{sec:methods}

Obtaining age and mass distributions of the VFTS stars in \tdor is a three stage process. First, the observed spectra are modelled with stellar atmosphere codes to determine spectroscopic parameters such as surface gravity, effective temperature and luminosity. Second, the spectroscopic parameters are matched against stellar evolutionary models to determine the ages and initial masses of each star. Third, the inferred ages and masses are combined to finally infer distributions of age and initial mass of various samples of the VFTS stars. The first two steps have been completed in \citet{2018Sci...359...69S} and we only briefly describe the most important aspects here.

Stars with composite spectra, for example due to binary stars or visual multiples, have been removed from our sample. Binaries are identified thanks to the multi-epoch nature of VFTS \citep{2013A&A...550A.107S,2015A&A...580A..93D} and visual multiples and nearby contaminating sources by comparison of the position of the \flames fibres on the sky with Hubble-Space-Telescope photometry \citep[GO12499, PI: D. Lennon;][]{2014A&A...564A..40W,2015A&A...580A..93D}. We further exclude those stars for which it is not possible to obtain reliable spectroscopic parameters or to find stellar models that reproduce all observables within their uncertainties simultaneously. Also, stars cooler than $9000\,\mathrm{K}$ have been removed from our sample because it is difficult to obtain robust ages and masses from their position in the Hertzsprung--Russell (HR) diagram alone. There are at most four such cool objects that are initially more massive than $15\,\msun$ and younger than $10\,\myr$. Their exclusion does therefore not affect our results and conclusions (see Sect.~\ref{sec:hrd}). 

We will now briefly describe how we determine stellar parameters (Sect.~\ref{sec:stellar-parameters}), age and mass distributions (Sect.~\ref{sec:inferring-age-mass-distributions}), and which completeness corrections have been applied to correct for biases introduced by our sample selection (Sect.~\ref{sec:completeness-corrections}). For more details and a full list of stellar parameters see \citet{2018Sci...359...69S} and Table~S3 therein.

\subsection{Determination of stellar parameters}\label{sec:stellar-parameters}

The observed VFTS spectra have been analysed in several steps, depending on spectral type and luminosity class. The hydrogen-rich Wolf--Rayet (WNh) and slash stars are modelled in \citet{2014A&A...570A..38B}, the O-type giants and supergiants in \citet{2017A&A...600A..81R}, the O-type dwarfs in \citet{2014A&A...564A..39S,2017A&A...601A..79S}, the B-type supergiants in \citet{2015A&A...575A..70M}, and the B-type dwarfs, classical Wolf--Rayet stars and cool-type stars (A-type and later) in \citet{2018Sci...359...69S}. For each star, we usually match the inferred effective temperature, surface gravity, luminosity and projected rotational velocity against the stellar models of \citet{2011A&A...530A.115B} and \citet{2015A&A...573A..71K} using the Bayesian code \bonnsai\footnote{The \bonnsai web-service is available at \url{http://www.astro.uni-bonn.de/stars/bonnsai}.} \citep{2014A&A...570A..66S,2017A&A...598A..60S} to determine full posterior probability distributions of age and initial mass. The above mentioned atmospheric parameters are not always all available and Table S3 in \citet{2018Sci...359...69S} contains the exact atmospheric parameters that have been used to match individual stars against the stellar models. We assume uniform prior distributions for age and initial mass, use the observed rotational velocity distributions of OB stars \citep{2013A&A...560A..29R,2013A&A...550A.109D} as initial rotational velocity prior distribution, and assume that rotational axes are randomly orientated in space. Post main-sequence (MS) stars are not covered by the stellar models of \citet{2011A&A...530A.115B} and \citet{2015A&A...573A..71K}, and the inference of their masses and ages requires special procedures \citep[see][for more details]{2018Sci...359...69S}. This holds true for so-called Hertzsprung--gap (HG) stars that left the main-sequence and evolve on a thermal timescale across the HR diagram to the red (super)giant branch, and for classical WR stars.

The present-day evolutionary masses determined by matching all available observables against stellar models are, on average, in agreement with the spectroscopic masses, $M_\mathrm{spec}=1/(4\pi\sigma G)\, (gL/T_\mathrm{eff}^4)$, derived directly from surface gravity $g$, effective temperature $T_\mathrm{eff}$ and luminosity $L$ (see Appendix~\ref{sec:mass-discrepancy}). This is a consistency check of our methods and shows that the inferred stellar parameters are on average robust.

\subsection{Determination of age and mass distributions}\label{sec:inferring-age-mass-distributions}

Once we have full posterior probability distributions of the ages and masses of each star, we sum them to obtain age and mass distributions of samples of stars. We apply completeness corrections (Sect.~\ref{sec:completeness-corrections}) to correct for biases introduced by our sample selection. Using a bootstrap technique, we estimate $1\sigma$ uncertainties of our age and mass distributions that allow us to judge the significance of individual features in the distributions. The uncertainties are the standard deviations of 10,000 realisations of age and mass distributions of stellar populations that have been randomly sampled with replacement, and thus mainly encompass information about the sample size and sample selection.

Our inferred age and mass distributions are not star formation histories (SFHs) and initial mass functions (IMFs), respectively, because our distributions do not account for those stars that already ended their nuclear burning lifetime. In \citet{2018Sci...359...69S}, we describe how to correct for such stars and infer the true SFH and IMF. Here, we use the same techniques to also infer the IMF in distinct sub-regions of \tdor.

\subsection{Completeness corrections}\label{sec:completeness-corrections}

The VFTS was designed to observe as many stars as possible with V-band magnitudes brighter than $V=17\,\mathrm{mag}$ with the \flames instrument \citep{2011A&A...530A.108E}. Because of the use of the \flames fibres, very crowded regions such as the core region around the R136 star cluster were avoided (see Fig.~\ref{fig:age-map-whole-region} below and, in particular, Fig.~\ref{fig:age-map-R136} for the position of VFTS targets near the crowded R136 star cluster), but there are no further biases. Currently, it is unknown how many massive stars have been missed in this crowded region and it is difficult to assess the completeness of the VFTS. Follow-up spectroscopy with the Hubble Space Telescope \citep{2016MNRAS.458..624C} revealed at least 57 O stars in the cluster core, within $0.5\,\mathrm{pc}$ of R136a1. Ongoing efforts with VLT MUSE will soon give a much clearer picture of the massive star content in the central region of NGC~2070 \citep{2018A&A...614A.147C,2018arXiv180100855C}. As shown in \citet{2018Sci...359...69S}, the completeness of the VFTS with respect to a more complete census of stars in \tdor \citep{2013A&A...558A.134D}, is independent of V-band magnitude, indicating an unbiased target selection. The spectroscopic completeness of the \citeauthor{2013A&A...558A.134D} census is estimated to be about 85\% outside $5\,\mathrm{pc}$ and about 35\% within the R136 region. However, our sample selection, for example against binaries, introduces biases that we need to correct for. As described in \citet{2018Sci...359...69S}, we apply four corrections when constructing age and mass distributions: Firstly, our sample selection introduce a completeness that varies with spectral types and luminosity class, which we correct for. Secondly, because crowded regions were avoided, we apply an averaged spatial completeness correction as a function of distance from the R136 cluster core that utilises the more complete census of stars in \tdor of \citet{2013A&A...558A.134D} as reference. Thirdly, a subset of stars in the VFTS were observed with the \argus instrument instead of \flames \citep{2011A&A...530A.108E}. However, only six emission line objects of the 37 VFTS \argus targets have yet been analysed, introducing a bias towards emission line stars which we correct for. Fourthly, the sample of \citet{2014A&A...570A..38B} contains the massive (${\approx}\,190\,\msun$) supergiant Mk\,42 which we remove from our sample because it has not been included in the VFTS. Including this object in our discussion might introduce a bias which we want to avoid. These four corrections hardly change our derived age and mass distributions, and hence our results and conclusions are robust (see Appendix~\ref{sec:am-no-completeness-correction}).

\section{Results}\label{sec:results}

\begin{figure*}
\begin{centering}
\includegraphics[width=0.90\textwidth]{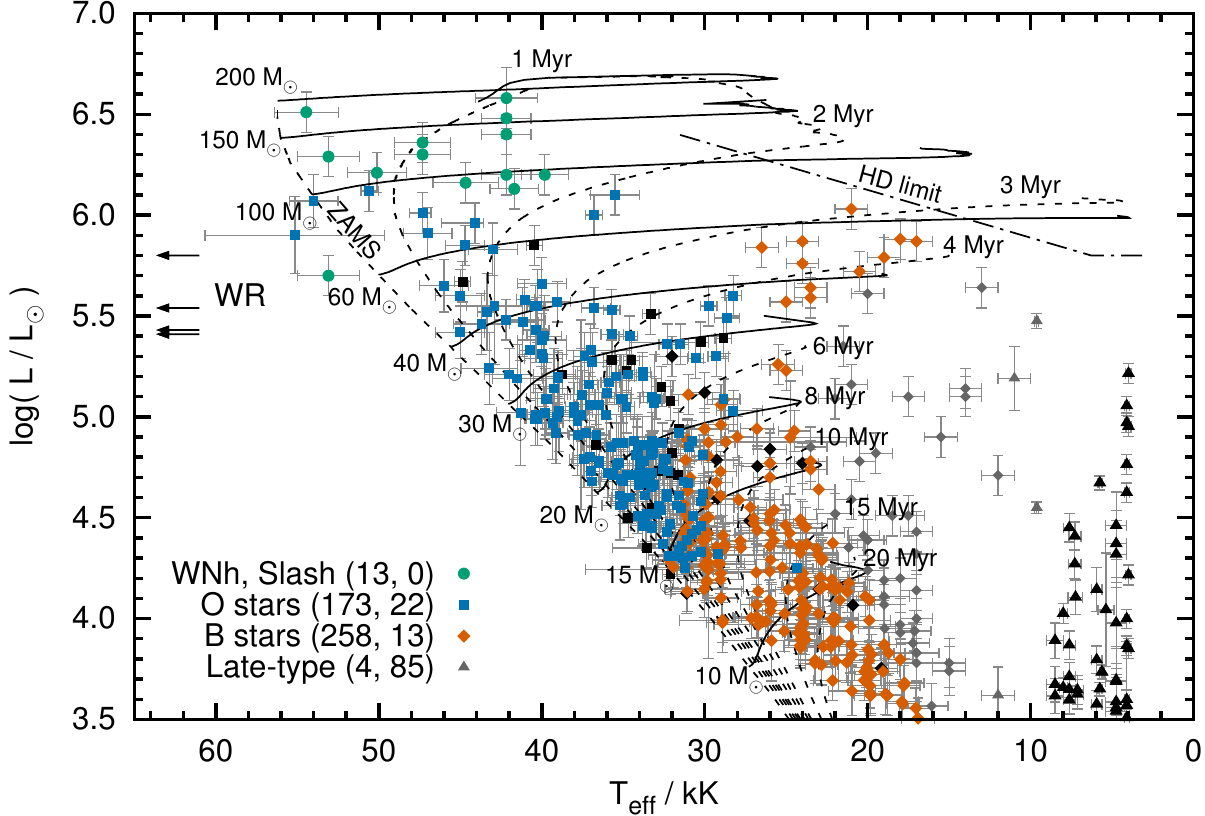}
\par\end{centering}
\caption{Hertzsprung--Russell diagram of the VFTS stars analysed in this paper. The solid and dashed black lines are the non-rotating stellar evolutionary tracks and isochrones of \citet{2011A&A...530A.115B} and \citet{2015A&A...573A..71K}, respectively. The Humphreys--Davidson (HD) limit is shown by the dot-dashed line in the top-right. Black symbols represent stars which cannot be reproduced by the stellar models (for reference, stars cooler than $9000\,\mathrm{K}$ and hence removed from our sample are also shown in black). In the legend, we provide the number of stars of each class that are in our final sample and those where not all observables can be reproduced by the stellar models simultaneously in parenthesis. Symbols in grey denote stars that fall off the MS part of the stellar tracks and are hence treated as Hertzsprung--gap stars (Sect.~\ref{sec:stellar-parameters}), but the symbols themselves keep their meaning. The luminosities of the four classical WR stars in our sample are indicated by the arrows on the left (they have surface temperatures of $80\text{--}90\,\mathrm{kK}$).}
\label{fig:hrd}
\end{figure*}

Overall, there are 934 stars in the VFTS of which 487 (52\%) are in our final sample (after removing identified binaries, multiple systems, stars with contaminated spectra etc.); of these 487 stars, 35 (7\%) cannot be reproduced by the used stellar models, leaving us with a sample of 452 stars with robust stellar parameters. Our final sample of 452 stars comprises 13 hydrogen-rich Wolf--Rayet and slash stars, four classical Wolf--Rayet stars, 173 O-stars, 258 B-stars and four A-stars. We first discuss our sample stars in the HR diagram with respect to stellar models (Sect.~\ref{sec:hrd}) before presenting their inferred age and mass distributions in spatially distinct regions of \tdor (Sect.~\ref{sec:age-mass-distributions}). We further quantify the spatial distribution of stellar ages and masses (Sect.~\ref{sec:spatial-distribution}), and derive the IMFs of massive stars in the different regions of \tdor (Sect.~\ref{sec:imf}).

\subsection{Hertzsprung--Russell diagram}\label{sec:hrd}

Our stellar sample populates the HR diagram with stars of all evolutionary stages, from very young stars near the zero-age main-sequence (ZAMS) to Hertzsprung--gap stars, red supergiants and classical WR stars (Fig.~\ref{fig:hrd}). The sample further fills the HR diagram densely with stars from less than $15$ up to about $200\,\msun$ and thus provides a unique opportunity to probe the evolution of massive stars across a wide range of masses and evolutionary stages. Since the VFTS sample is magnitude limited, we find a temperature-dependent cut-off in the HR diagram at effective temperatures $T_\mathrm{eff}\,{\lesssim 30}\,\,\mathrm{kK}$ and luminosities $\log L/\lsun\, {\lesssim}\, 4.0$.

The empirical ZAMS seems to coincide with that of the models but we note that our sample lacks stars younger than ${\approx}\,1\,\myr$ (Sect.~\ref{sec:overall-sf-process}). This lack results in a sparsely populated region in the HR diagram between the ZAMS, the $1\,\myr$ isochrone and the $30$ and $100\,\msun$ stellar tracks. 

There is a clear dearth of stars at effective temperatures of about $18\text{--}9\,\mathrm{kK}$ that seems to extend up to the Humphreys--Davidson (HD) limit \citep{1979ApJ...232..409H} at around $\log L/\lsun\, {\approx}\, 5.9$. This dearth of stars is where stellar models predict the so-called Hertzsprung--gap that, in the past, was often found to be densely populated with blue supergiants, giving rise to the blue-supergiant problem \citep[\eg][]{1989AJ.....98.1598B,1990ApJ...363..119F}. The VFTS sample is unbiased for stars with initial masses ${\geq}\, 15\,\msun$, implying that the lower star density around $T_\mathrm{eff}\,{\approx}\, 15\,\mathrm{kK}$ and $\log L/\lsun\, {\gtrsim}\, 4.8$ compared to the higher star density towards the MS is not due to a selection bias. The star-formation rate in \tdor accelerated $8\text{--}10\,\myr$ ago \citep{2018Sci...359...69S}, reducing the number densities on the cooler and lower luminosity side of the $8\text{--}10\,\myr$ isochrone in the HR diagram. For stars less massive than $15\,\msun$, our sample becomes incomplete such that the number densities are not a reliable indicator of stellar physics or star formation. However, even in this regime, we do find a gap of stars at $T_\mathrm{eff}\, {\approx}\, 13\,\mathrm{kK}$ in the HR diagram that separates the red supergiant branch from the main sequence. A gap in the SFH would produce a reduced stellar density along an isochrone, but not such a gap. It therefore seems that the blue-supergiant problem and the lack of a Hertzsprung--gap are less severe with the current data and stellar models.

The terminal-age main-sequence (TAMS) of the models does not coincide with the beginning of the Hertzsprung--gap but seems to be slightly hotter. Convective core overshooting plays an important role in determining the position of the TAMS and our data may suggest that the models require more overshooting than is currently used \citep[see also discussion in the VFTS predecessor survey, \eg][]{2010A&A...512L...7V}. This idea is further supported by some VFTS B supergiants that are in relatively close binaries (such that binary mass transfer has most likely not occurred yet) and that are apparently cooler than the TAMS of the stellar models \citep[see][]{2015A&A...575A..70M}. There are further ways to push the apparent TAMS to cooler temperatures for example by merger products \citep[\eg][]{2014ApJ...782....7D,2014ApJ...796..121J} and uncertain physics such as semi-convection \citep[\eg][]{1989A&A...224L..17L,1991A&A...252..669L}. The consequences of our finding therefore need to be worked out more carefully but is---in general---in line with the work of \citet{2014A&A...570L..13C} who suggest extra overshooting in more massive stars from an empirical TAMS in the spectroscopic HR diagram of Galactic OB stars. At lower masses (${\lesssim}\,10\,\msun$), there is also evidence for mass-dependent core overshooting \citep[\eg][]{2000MNRAS.318L..55R,2016A&A...589A..93D,2016A&A...592A..15C} but these results are not unchallenged \citep[\eg][]{2014ApJ...787..127M,2015A&A...575A.117S}.

In the models of \citet{2011A&A...530A.115B} and \citet{2015A&A...573A..71K}, stars more massive than about $50\,\msun$ develop inflated envelopes as a consequence of reaching the Eddington limit in their envelopes because of iron opacity bumps \citep{2015A&A...580A..20S}. Inflation pushes the TAMS to cooler temperatures and may help explain the positions of some blue supergiants. Comparing model predictions (\eg figs.~2 and~10 in \citealt{2015A&A...580A..20S}) and the observations, inflation may be relevant for the group of B-type supergiants at $T_\mathrm{eff}\, {\approx}\, 22\,\mathrm{kK}$ and $\log L/\lsun\, {\approx}\, 5.8$.

Two stars, VFTS~108 (WN7h) and~125 (Ope), appear to be hotter than the ZAMS. Chemically homogeneously evolving models can explain their observables well. If true, these two stars are expected to rotate relatively fast (present-day rotational velocities of ${\gtrsim}\, 170\,\kms$ and ${\gtrsim}\, 250\,\kms$, respectively) and may retain enough angular momentum to produce long-duration gamma-ray-bursts at the end of their lives \citep[\eg][]{2005A&A...443..643Y,2006ApJ...637..914W}. Further candidates of this evolutionary channel have been identified \citep{2009A&A...495..257M,2013A&A...554A..23M,2015ApJ...812..102A}. There is the alternative possibility that VFTS~108 and~125 are classical WR stars with highly inflated envelopes that place the stars so close to the ZAMS as discussed by \citet{2014A&A...570A..38B}. The large projected rotational velocity of VFTS~125 of about $270\,\kms$ is rather unexpected for a classical WR star with an inflated envelope and thus seems to favour a chemically homogeneous star. The projected rotational velocity of VFTS~108 is not well measured (it seems to be smaller than $200\,\kms$) and is thus of limited help for constraining its evolutionary status. We note that the effective temperature of VFTS~125 is considerably uncertain and it can therefore not be excluded that it is actually cooler than the ZAMS and thus not a strong candidate of a chemically homogeneously evolving star.

We find no star above the HD limit. Furthermore, no star is found to be cooler than $35\,\mathrm{kK}$ and brighter than $\log L/\lsun\, {=}\, 6.1$ despite the fact that initially slowly rotating (${\lesssim}\, 100\,\kms$), $100\text{--}200\,\msun$ stellar models of \citet{2015A&A...573A..71K} spend about $30\text{--}40\%$ of their MS lifetime in this part of the HR diagram. Given that there are 14 stars hotter than $35\,\mathrm{kK}$ and brighter than $\log L/\lsun\, {=}\, 6.1$, the models predict five to six stars in a region of the HR diagram where no star is observed, a more than $2\sigma$ mismatch\footnote{The implicit assumption here is that stars formed continuously and at a similar rate over the last $2\text{--}3\,\myr$, which seems reasonable \citep[\eg][]{2018Sci...359...69S}.}. 

Another independent hint towards a mismatch of the massive star models and the observed stars comes from the surface helium enrichment of stars brighter than $\log L/\lsun\, {=}\, 6.1$ \citep[for helium abundances see Table~S3 in][]{2018Sci...359...69S}. Surface helium enrichment is obtained in two ways in single star models: first, by the removal of the hydrogen envelope through stellar winds which exposes material processed by nuclear burning and, second, by (rotationally induced) mixing that transports chemically enriched material from the core to the surface. An initial rotational velocity of larger than $300\,\kms$ is required in eight to nine of these 14 bright stars, that is in more than $50\text{--}60\%$ of them, to explain the high surface helium mass fractions, because the stellar wind in our models is---on its own---not efficient enough to enrich the stellar surfaces to the observed values. We consider such a large fraction of initially fast rotators implausible given that less than 20\% of OB stars in \tdor rotate that fast \citep{2013A&A...550A.109D,2013A&A...560A..29R}. Hence, single star models alone likely lack physics to fully describe the upper part of the HR diagram.

\begin{figure*}
\begin{centering}
\includegraphics[width=0.95\textwidth]{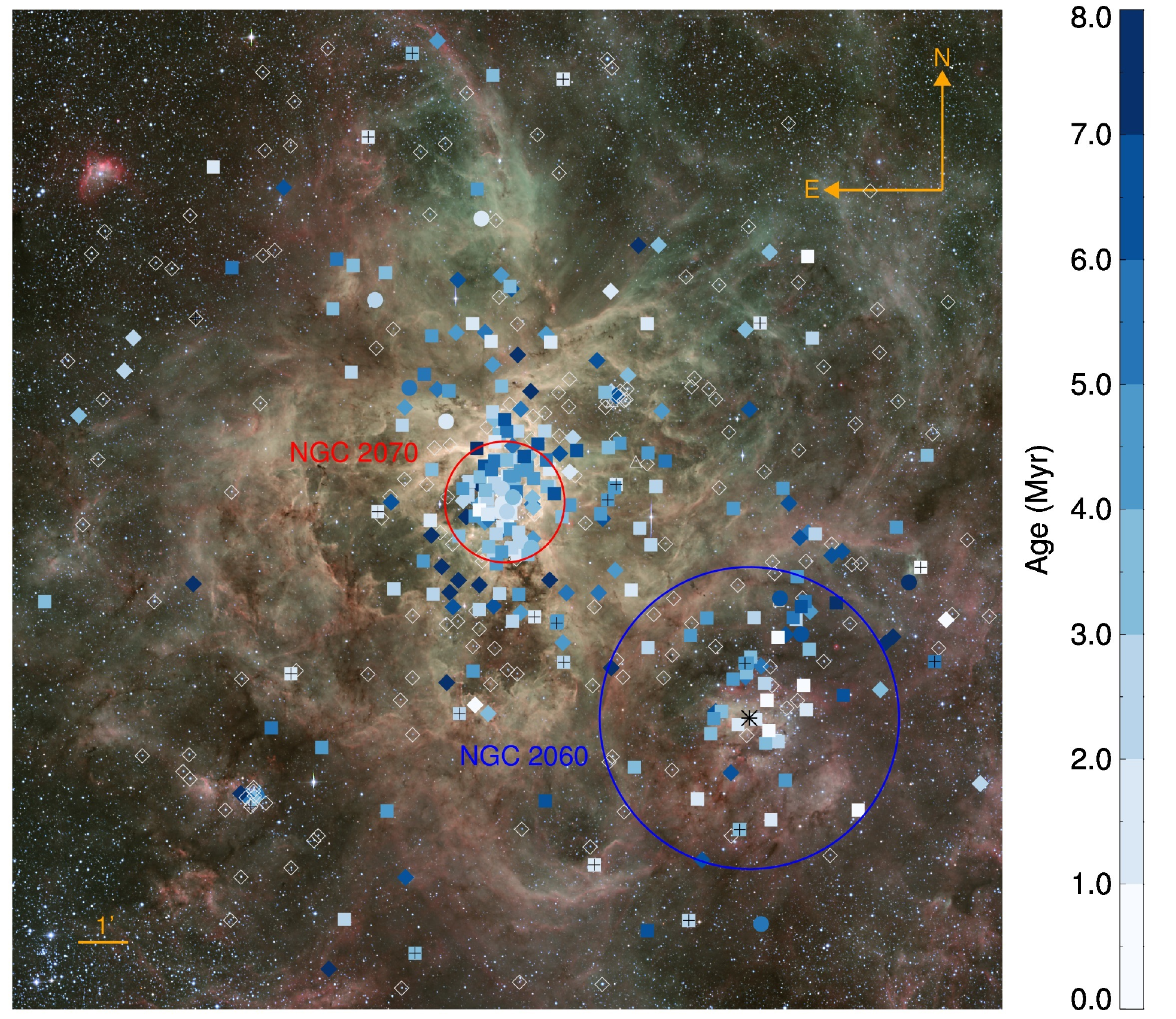}
\par\end{centering}
\caption{Age map of our sample of massive VFTS stars. Symbols are as in Fig.~\ref{fig:hrd}, \ie circles denote slash, WNh and WR stars, squares O stars, diamonds B stars and triangles later-type stars. Runaway stars are marked by additional plus-signs and the position of the pulsar \pulsar by a black asterisk. The red and blue circles show regions of $1.2\,\mathrm{arcmin}$ (${\approx}\,18\,\mathrm{pc}$) around R136 (NGC~2070) and $3\,\mathrm{arcmin}$ (${\approx}\,44\,\mathrm{pc}$) around the pulsar (NGC~2060), respectively. Symbols without filling indicate stars older than $8\,\myr$. The background image of 30~Dor is an optical (B, V, [\ion{O}{iii}] and H$\alpha$) composite taken with the Wide Field Imager (WFI) at the MPG/ESO 2.2-m telescope on La Silla under programme ID 076.C-0888, processed and released by the ESO VOS/ADP group.}
\label{fig:age-map-whole-region}
\end{figure*}

\citet{2014A&A...570A..38B} come to a similar conclusion when studying the mass-loss properties of the most massive stars in \tdor and find evidence for a wind enhancement of stars close to the Eddington limit as theoretically predicted by for example \citet{2008A&A...482..945G} and \citet{2011A&A...531A.132V}. An increased wind mass loss with Eddington factor may help to remove the tension between the stellar models and observations in this part of the HR diagram. Additionally, some of the very massive stars may be products of binary mass exchange \citep[\eg][]{2014ApJ...780..117S,2014ApJ...782....7D}, that could explain their large observed surface helium abundances and hotter temperatures.

There are three stars in our sample (VFTS~682, 1022 and~1025) with an initial mass in excess of $150\,\msun$. In the past, this mass has been suggested to be the largest possible birth mass of stars \citep[\eg][]{2004MNRAS.348..187W,2005Natur.434..192F,2005ApJ...620L..43O,2006MNRAS.365..590K}. VFTS~1025 aka R136c belongs to the four very massive stars in the R136 star cluster already found by \citet{2010MNRAS.408..731C} to exceed this limit. There are probably at least two further stars in \tdor that are initially more massive than $150\,\msun$: Mk\,42 and VFTS~695. The latter likely has a composite spectrum and is therefore removed from our sample, and Mk\,42 is not part of the VFTS sample and is hence not considered in this work to avoid modifying our selection strategy (see Sect.~\ref{sec:completeness-corrections}). So there are currently about eight candidates with initial masses in excess of $150\,\msun$. Some of them may be merger products \citep[\eg][]{1999A&A...348..117P,2012MNRAS.426.1416B,2014ApJ...780..117S} and further work is needed to establish the origin of their high masses \citep[see also discussion in][]{2015HiA....16...51V}.

\subsection{Age and mass distributions of massive stars in \tdor}\label{sec:age-mass-distributions}

\subsubsection{Full \tdor region}\label{sec:am-whole-region}

\begin{figure*}
\begin{centering}
\includegraphics[width=0.95\textwidth]{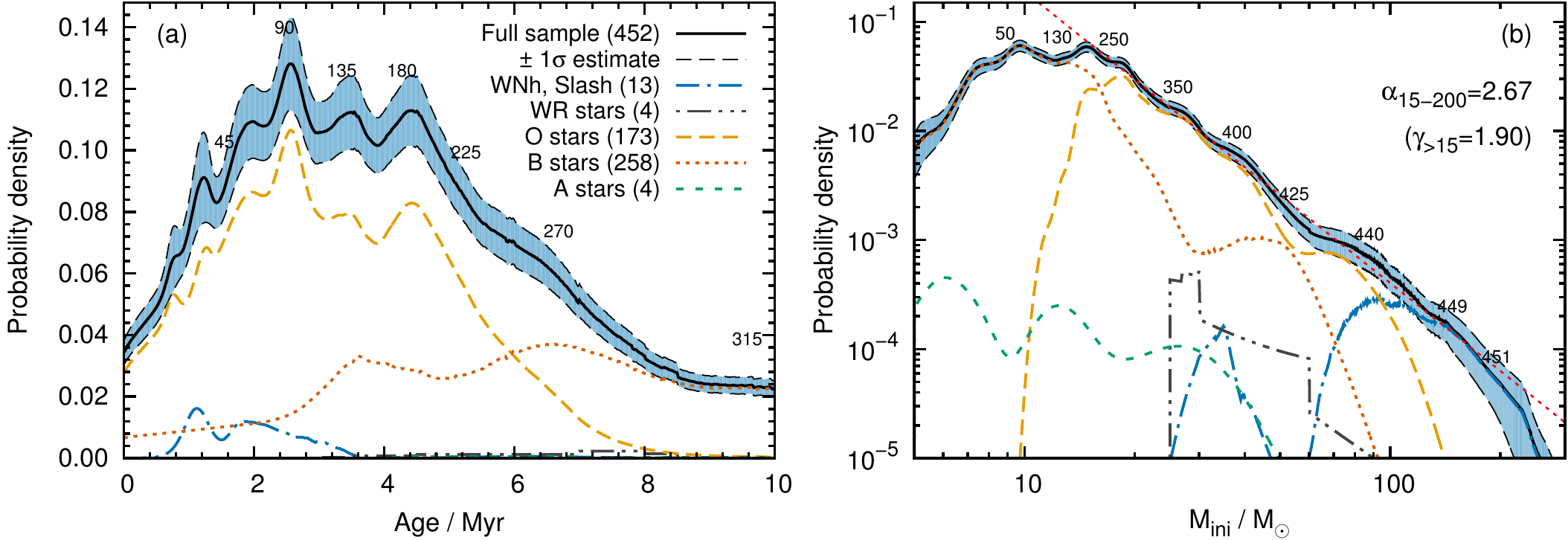}
\par\end{centering}
\caption{Probability density functions of (a) ages and (b) initial masses of all stars in our sample. The contributions of slash and WNh, WR, O, B and A stars are shown. The blue shaded areas are bootstrapped $\pm1\sigma$ estimates (Sect.~\ref{sec:inferring-age-mass-distributions}) and the numbers along the curves indicate the cumulative number of stars up to certain ages and masses to judge which features of the distributions are populated by how many stars. The age distribution beyond $10\,\myr$ keeps dropping and is essentially zero after ${\approx}\,40\,\myr$. In panel (b), a power-law mass-function, $\zeta(M)=\mathrm{d}p/\mathrm{d}M \propto M^{-\alpha}$, is fitted to the mass distribution and the inferred high-mass IMF slope $\gamma_{>15}$ after correcting for those stars that already ended their nuclear burning lifetime is also provided (see Sect.~\ref{sec:imf}). For reference, a Salpeter IMF has a power-law slope of $\gamma\, {=}\, 2.35$ \citep{1955ApJ...121..161S}.}
\label{fig:age-mass-distr-whole-region}
\end{figure*}

In Fig.~\ref{fig:age-map-whole-region}, we show the positions of our sample stars in 30~Dor and colour-code their ages. Massive stars cluster around R136 in the NGC~2070 region encircled in red and the NGC~2060 region encircled in blue. The radii of the two circles are chosen arbitrarily such that they roughly contain the same number of VFTS stars. The energetic $16\,\mathrm{ms}$ X-ray pulsar \pulsar and its supernovae remnant N157B reside in the centre of NGC~2060 \citep{2006ApJ...651..237C}. Hodge~301, a $15\text{--}30\,\myr$ old star cluster \citep{2000AJ....119..787G,2015A&A...574A..13E,2016ApJ...833..154C} hosting three red supergiants and VFTS B stars, lies about $3\,\mathrm{arcmin}$ ($44\,\mathrm{pc}$) north-west of R136, and the star cluster SL~639 \citep[$10\text{--}15\,\myr$;][]{2015A&A...574A..13E} about $7.5\,\mathrm{arcmin}$ ($110\,\mathrm{pc}$) south-east of R136. 

Stars of all ages are scattered throughout the \tdor region and there are no obvious spatially coherent age patterns. Wherever groups of massive stars are found, they span a considerable range of ages. This finding already provides important information regarding the formation process of massive stars: either stars formed rather randomly in \tdor with some high concentrations of stars such as in and around NGC~2070, or they formed in dense star clusters and migrated in some cases over large distances (tens to hundred of pc) to be found at their current positions in such a scattered fashion. We will come back to this aspect later.

In order to study the star formation process more quantitatively, we show the distribution of ages and initial masses in Fig.~\ref{fig:age-mass-distr-whole-region}. The age distribution starts to increase at about $8\,\myr$ and reaches a plateau at $1.5\text{--}5.0\,\myr$ before it declines again; the median age of our full stellar sample is $5.3\,\myr$. We only show the age distributions up to $10\,\myr$, because there are no features beyond this age. The age distribution simply levels off and essentially drops to zero by about $40\,\myr$.

There are wiggles, peaks and shoulders in the age distribution in Fig.~\ref{fig:age-mass-distr-whole-region}a (for example at ${\approx}\,2.5$, ${\approx}\,4.4$ and ${\approx}\,6.4\,\myr$) that are not significant given the bootstrapped error estimates. Given the current accuracy and precision of the stellar ages, we are unable to easily identify coeval groups of stars. However, some features may indicate groups of stars that formed at the same time either spatially localised in for example star clusters or stellar associations, or non-localised during an enhanced star formation period in \tdor. We will show below (Sects.~\ref{sec:am-r136-region}--\ref{sec:am-stars-outside}) that some of the features in Fig.~\ref{fig:age-mass-distr-whole-region}a can be attributed to certain regions in \tdor.

The distribution of initial masses reaches a plateau below about $15\,\msun$, indicating the mass threshold above which our sample is unbiased. Also the HR diagram (Fig.~\ref{fig:hrd}) indicates the same $15\,\msun$ threshold. At this mass, the VFTS magnitude limit does not affect the ZAMS. The mass distribution is well sampled up to at least $200\,\msun$ and can be fitted with a single power-law ($\zeta(M)=\mathrm{d}p/\mathrm{d}M \propto M^{-\alpha}$) of slope $\alpha{=}2.67$ in the mass range $15\text{--}200\,\msun$. The slope is steeper than the $2.35$ of the Salpeter IMF \citep{1955ApJ...121..161S} because our mass distribution represents the distribution of initial masses of stars observed today, meaning that it lacks those stars that already ended their nuclear burning lifetime. In Sect.~\ref{sec:imf}, we infer the true IMF slopes of our sample stars from the age and mass distributions in Fig.~\ref{fig:age-mass-distr-whole-region}.

Our sample is only unbiased above $15\,\msun$ and from the VFTS magnitude limit in the HR diagram (Fig.~\ref{fig:hrd}) it can be seen that we lack young B and A stars that would be located close to the ZAMS. This implies that our age and mass distributions of B and A stars are biased towards older and more massive stars. Quantitative results derived from these distributions may therefore be used within this study for differential comparisons, but should not be compared to other studies where stars have been selected differently.

There are ten B stars and one O star within $1/3\,\mathrm{arcmin}$ of Hodge~301, and twelve B stars within $1/3\,\mathrm{arcmin}$ of SL~639 in our sample that have median ages of $13.5\,\myr$ and $11.1\,\myr$, respectively. In all cases, it is not certain which of these stars are genuine cluster members and whether some are rejuvenated binary products, but these ages appear to be consistent with ages derived for these clusters in the literature \citep[\eg][]{2000AJ....119..787G,2015A&A...574A..13E}. However, we note that \citet{2016ApJ...833..154C} derive an older age of $25\text{--}30\,\myr$ for Hodge~301, which is at odds with the median age of VFTS sources in the vicinity of this cluster.

In the following, we investigate the age and mass distributions of our sample stars in three spatially distinct regions, namely in NGC~2070 (R136 region), in NGC~2060 and outside of these two regions to further disentangle and understand the age distribution in Fig.~\ref{fig:age-mass-distr-whole-region}a and the star formation process in \tdor.

\subsubsection{NGC~2070 (region around R136)}\label{sec:am-r136-region}

\begin{figure*}
\begin{centering}
\includegraphics[width=0.85\textwidth]{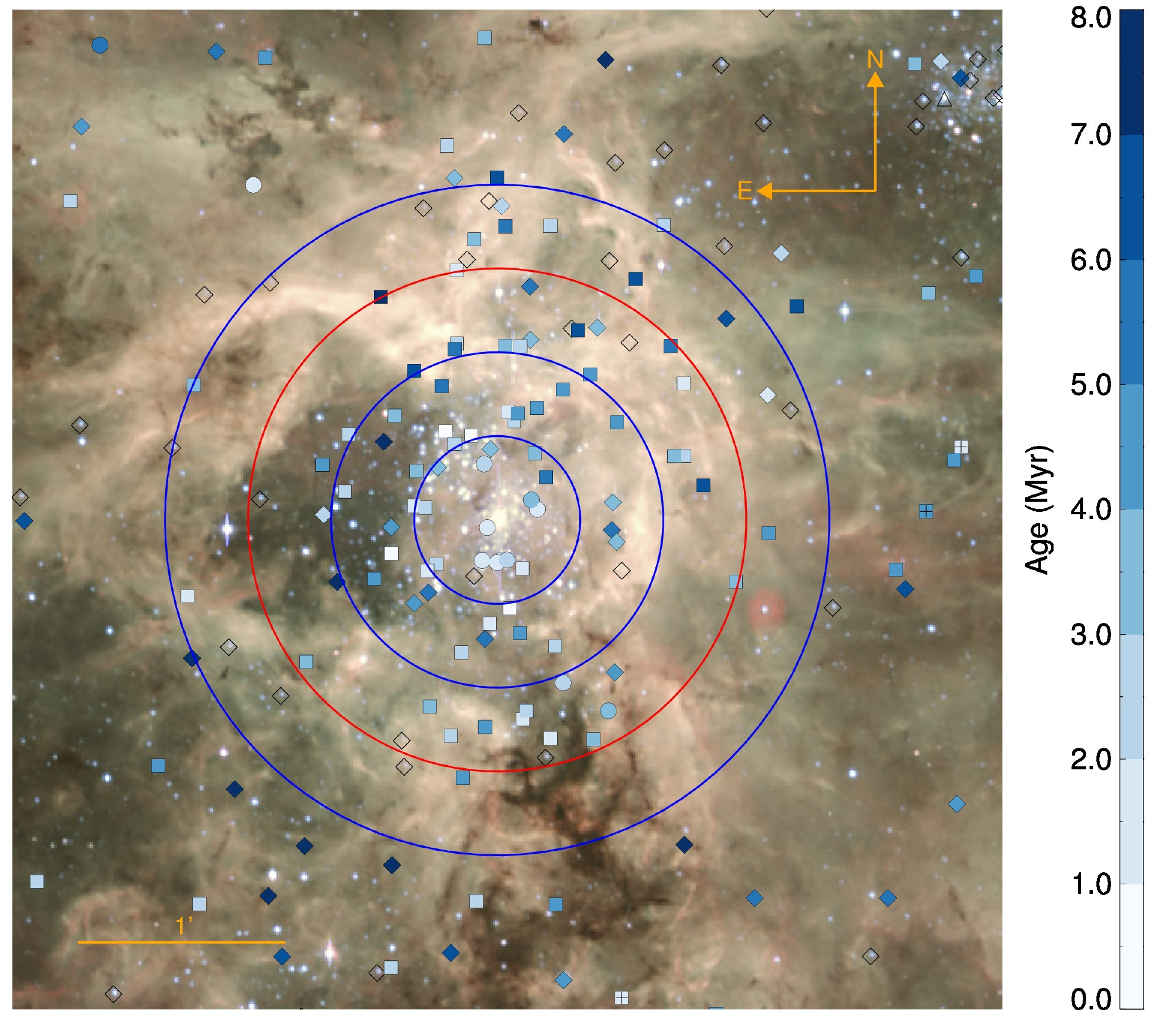}
\par\end{centering}
\caption{As Fig.~\ref{fig:age-map-whole-region} but zoomed into the NGC~2070 region around the R136 star cluster. To guide the eye, we show four circles centred on R136 with radii of $0.4$ ($6$), $0.8$ ($12$), $1.2$ ($18$) and $1.6\,\mathrm{arcmin}$ ($23\,\mathrm{pc}$). As is evident from the distribution of stars, the innermost region around the crowded R136 cluster has not been observed in the VFTS.}
\label{fig:age-map-R136}
\end{figure*}

\begin{figure*}
\begin{centering}
\includegraphics[width=0.95\textwidth]{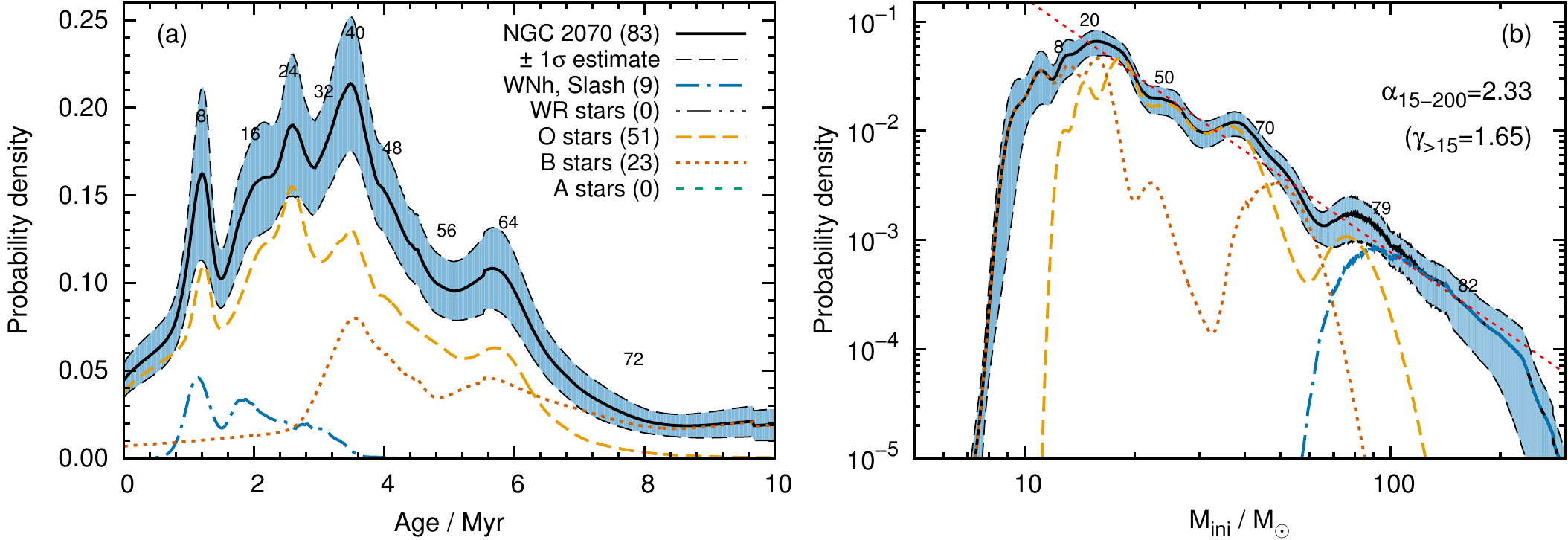}
\par\end{centering}
\caption{As Fig.~\ref{fig:age-mass-distr-whole-region} but for stars within $1.2\,\mathrm{arcmin}$ of R136 ($18\,\mathrm{pc}$; red circles in Figs.~\ref{fig:age-map-whole-region} and~\ref{fig:age-map-R136}).}
\label{fig:age-mass-distr-r136}
\end{figure*}

NGC~2070 resides at the heart of \tdor and is the most prominent site of recent, massive star formation. The R136 star cluster lies just in the centre of this region and we show the positions and ages of our sample stars around it in Fig.~\ref{fig:age-map-R136}. The core region of R136 has not been observed within the VFTS because of crowding and it is not immediately evident which of our sample stars are genuine R136 cluster members. This is in part also because we are observing the projection of a 3D structure on the sky which makes it difficult to assign an age to R136 from nearby stars in our sample.

The massive and young WNh stars VFTS~1001 and VFTS~1025 aka R136c are close to the R136 cluster core and may thus belong to R136. From the VFTS data, we have inferred ages of $3.0\pm0.3\,\myr$ and $1.8\pm0.2\,\myr$ for VFTS~1001 and VFTS~1025, respectively. To put a lower limit on the ages of these two stars, and thus to the age of R136, we consider their observed surface helium mass fractions. The surface helium mass fraction holds information of how much helium has at least been synthesised in stellar cores. Lower age limits can then be derived by assuming that the helium core mass fraction, $Y_\mathrm{core}$, is the same as that observed on the surface and that the core helium mass fraction scales linearly with age, that is $t/\tau_{\mathrm{MS}}=\left(Y_{\mathrm{core}}-Y_{\mathrm{ini}}\right)/\left(1-Y_{\mathrm{ini}}\right)$ with $t$ being the age of the star, $Y_{\mathrm{ini}}\,{\approx}\,0.26$ the initial helium mass fraction and $\tau_{\mathrm{MS}}$ the main-sequence lifetime. These age limits are larger if the star was initially rotating faster, prolonging the main-sequence life, or if the star is a binary product in which case it was likely rejuvenated. The surface helium mass fractions inferred by \citet{2014A&A...570A..38B} for VFTS~1001 and VFTS~1025 are $0.85\,{\pm}\,0.05$ and $0.70\,{\pm}\,0.05$, respectively. Given the observed luminosity of $\log L/\lsun\,{\approx}\,6.2$ and effective temperature of $T_{\mathrm{eff}}\,{\approx}\,42.2\,\mathrm{kK}$ of VFTS~1001, it falls on an initially $100\,\msun$ stellar track. Such stars have main-sequence lifetimes of about $2.6\,\myr$, meaning VFTS~1001 must be older than $1.7\,\myr$ with $98\%$ confidence. The confidence levels are derived from the uncertainties in the observed surface helium mass fraction. VFTS~1025, with an observed luminosity of $\log L/\lsun\,{\approx}\,6.6$ and effective temperature of $T_{\mathrm{eff}}\,{\approx}\,42.2\,\mathrm{kK}$, is close to an initially $200\,\msun$ track that has a main-sequence lifetime of about $2.1\,\myr$. Hence, VFTS~1025 is older than $1.0\,\myr$ with $98\%$ confidence. From our data, it therefore seems that stars in R136 are older than $1\,\myr$. \citet{2016MNRAS.458..624C} find a most likely cluster age for R136 of $1.5^{+0.5}_{-0.7}\,\myr$ (\ie $0.8\text{--}2.0\,\myr$). This is in agreement with our conclusion from VFTS~1001 and VFTS~1025.

The age and mass distributions of our sample stars within $1.2\,\mathrm{arcmin}$ of R136 (red circles in Figs.~\ref{fig:age-map-whole-region} and~\ref{fig:age-map-R136}) are shown in Fig.~\ref{fig:age-mass-distr-r136}. The median age is $3.6\,\myr$, meaning it is younger than that of the full \tdor region. Stars closer to R136 are, on average, younger than stars further out, giving rise to a core-halo age gradient: stars in the $0.0\text{--}0.4\,\mathrm{arcmin}$ annulus have a median age of $3.4\,\myr$, which increases to $3.6$, $4.4$ and $6.7\,\myr$ in the $0.4\text{--}0.8$, $0.8\text{--}1.2$ and $1.2\text{--}1.6\,\mathrm{arcmin}$ annuli, respectively. 

The age distribution in Fig.~\ref{fig:age-mass-distr-r136}a shows a (potential double) peak at $2\text{--}4\,\myr$ which represents the characteristic age of our sample stars in NGC~2070. This includes stars just north-east of R136, the north--east (NE) clump, that might be in the process of merging with R136 as suggested by \citet{2012ApJ...754L..37S}. The same authors find ages of $2\text{--}5\,\myr$ for pre-MS stars in the NE clump, which appears consistent with our age estimates of massive stars in this part of NGC~2070. A tail of about $6\,\myr$ old stars are primarily found ${>}\,0.4\,\mathrm{arcmin}$ from R136, contributing to the appearance of the core--halo age gradient.

The initial-mass distribution (Fig.~\ref{fig:age-mass-distr-r136}b) is filled up to ${\approx}\, 200\,\msun$ and the most massive and youngest stars are concentrated within $0.4\,\mathrm{arcmin}$ ($6\,\mathrm{pc}$) of R136. A power-law fit to the mass distribution gives a slope of $2.33$---we discuss the true IMF of this region in Sect.~\ref{sec:imf}.

\begin{figure*}
\begin{centering}
\includegraphics[width=0.85\textwidth]{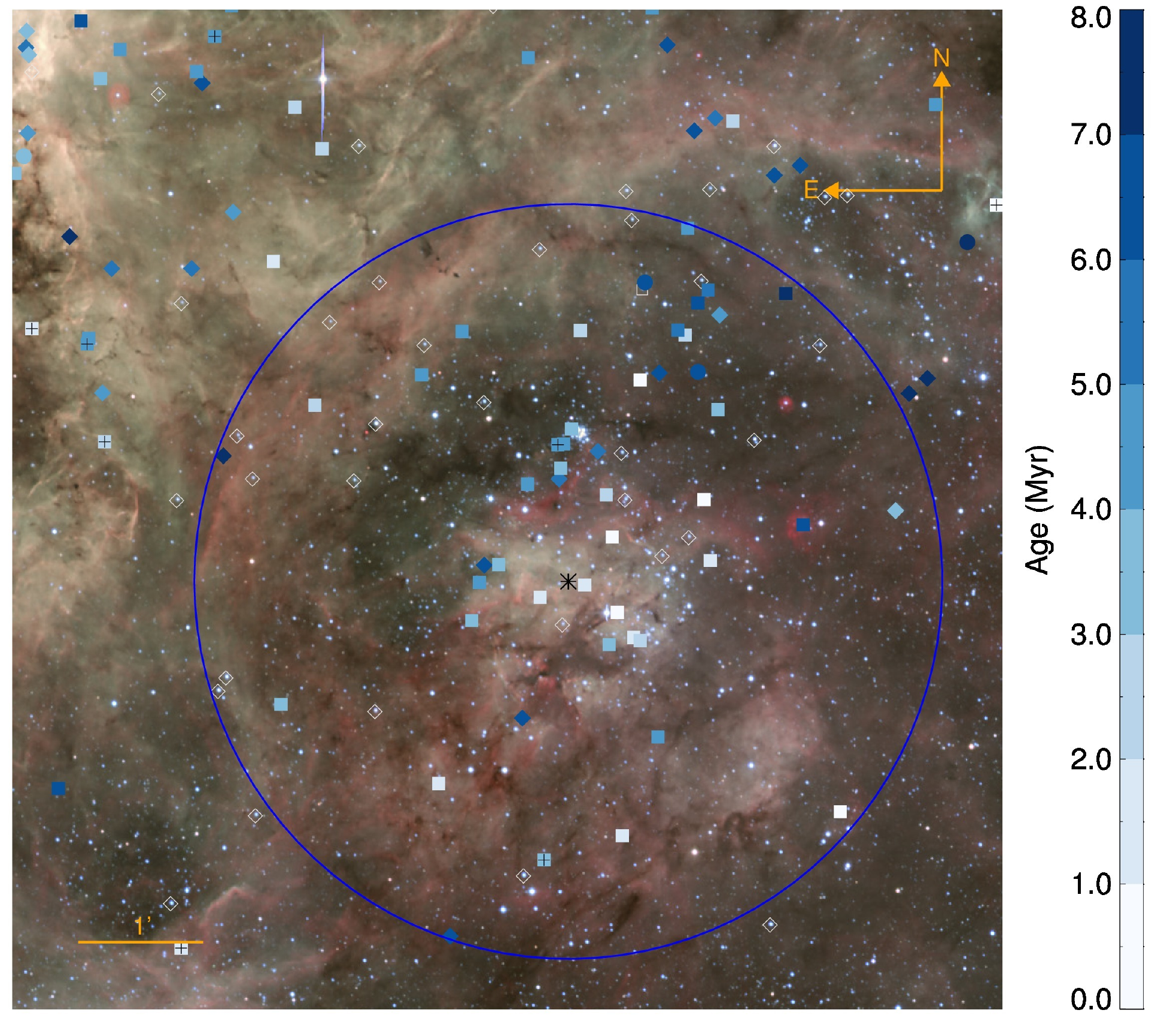}
\par\end{centering}
\caption{As Fig.~\ref{fig:age-map-whole-region} but zoomed into the NGC~2060 region. The blue circle of radius $3.0\,\mathrm{arcmin}$ ($44\,\mathrm{pc}$) is centred on the pulsar \pulsar (black asterisk) and stars within this circle are used to compute age and mass distributions.}
\label{fig:age-map-ngc2060}
\end{figure*}

\begin{figure*}
\begin{centering}
\includegraphics[width=0.95\textwidth]{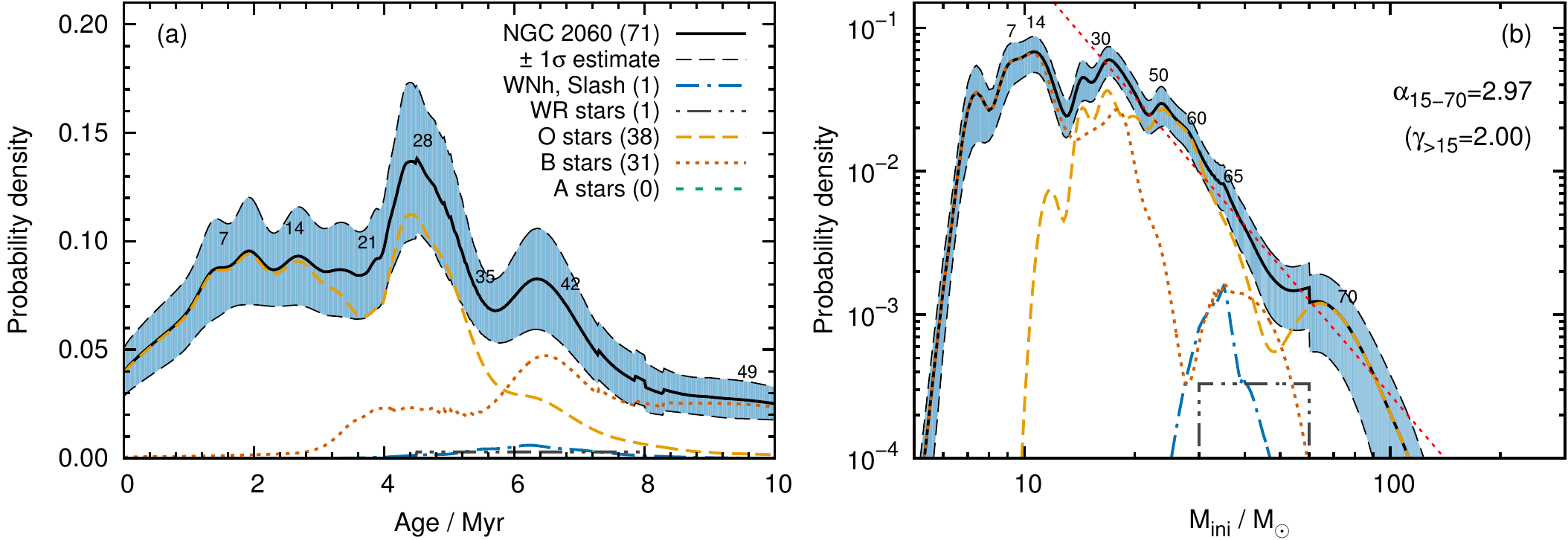}
\par\end{centering}
\caption{As Fig.~\ref{fig:age-mass-distr-whole-region} but for stars within $3.0\,\mathrm{arcmin}$ ($44\,\mathrm{pc}$) of the pulsar \pulsar in NGC~2060 (blue circles in Figs.~\ref{fig:age-map-whole-region} and~\ref{fig:age-map-ngc2060}).}
\label{fig:age-mass-distr-ngc2060}
\end{figure*}

\subsubsection{NGC~2060}\label{sec:am-ngc2060-region}

The second-highest concentration of OB stars is found in NGC~2060. In Fig.~\ref{fig:age-map-ngc2060} we show the positions and ages of our stars in this part of \tdor. The $16\,\mathrm{ms}$ pulsar \pulsar forms the centre of a $3\,\mathrm{arcmin}$ (${\approx}\,44\,\mathrm{pc}$) region that we use to compute the age and mass distributions of this sample (see Fig.~\ref{fig:age-mass-distr-ngc2060}). 

\begin{figure*}
\begin{centering}
\includegraphics[width=0.95\textwidth]{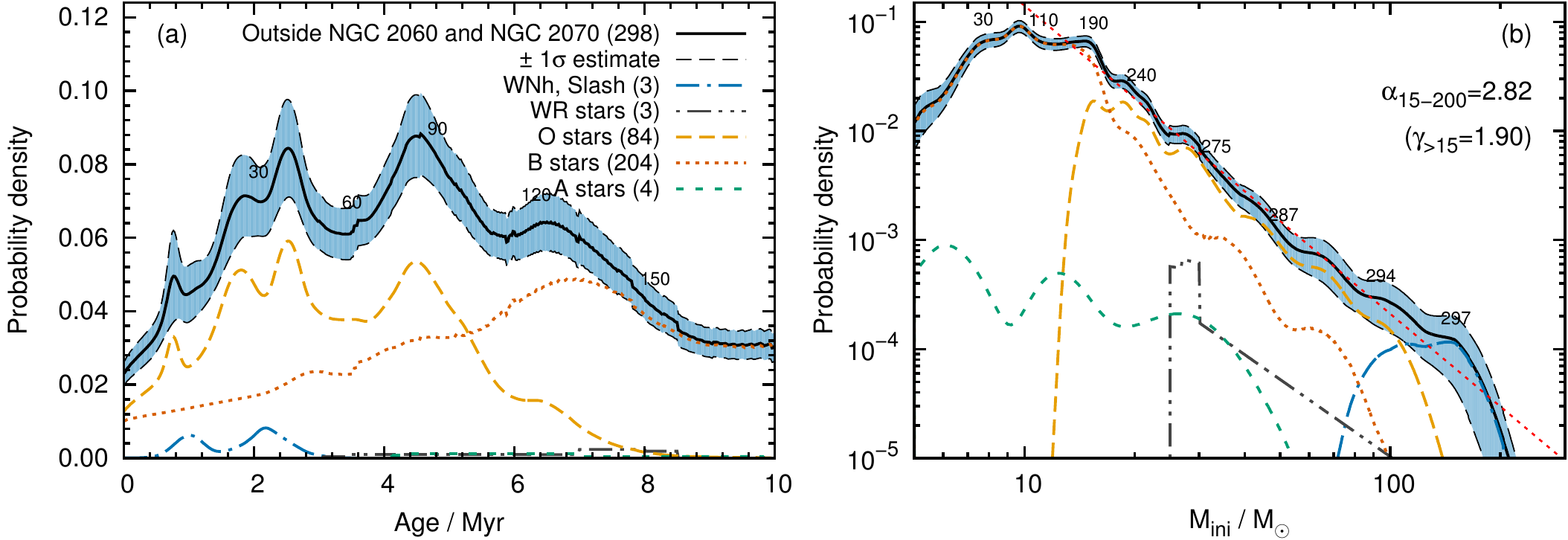}
\par\end{centering}
\caption{As Fig.~\ref{fig:age-mass-distr-whole-region} but for stars outside NGC~2060 and NGC~2070.}
\label{fig:age-mass-distr-outside}
\end{figure*}

About $1.2\,\mathrm{arcmin}$ north of the pulsar \pulsar lies the stellar association \tld (Brey~73) that hosts several OB stars and the WN6(h) star VFTS~147 \citep{1988Msngr..54...43T,1999AJ....117..238B,2008MNRAS.389..678B}. Within the \tld association are the O dwarf VFTS~154 and the O supergiants VFTS~141, 151 and~153. The latter two are visual multiples of at least five and three stars, respectively \citep{2014A&A...564A..40W}; the stellar models cannot reproduce the properties of VFTS~141 and the inferred stellar parameters of VFTS~147 are discarded as highly uncertain by \citet{2014A&A...570A..38B}. We thus only have one good age estimate for the \tld association from our stellar sample of $3.4^{+0.3}_{-0.3}\,\myr$ based on the age of VFTS~154.

Based on the spectral classifications of \citet{1999AJ....118.1684W} and the calibrations of \citet{2016MNRAS.458..624C} applied to stars in the R136 cluster core, we derive effective temperatures and luminosities for eight OB stars in \tld (Appendix~\ref{sec:tld1}). We determine masses and ages of these stars with the same methods and assumptions as described in Sect.~\ref{sec:stellar-parameters} and compute an age distribution (Fig.~\ref{fig:age-distr-tld1}). The age distribution peaks at about $3.3\,\myr$ (median age of $3.5\,\myr$) which is in good agreement with the age inferred for \tld from VFTS~154.

There seems to be an age gradient in NGC~2060 with the on average oldest stars located north to north-west of \pulsar and the on average youngest in its direct vicinity. The association \tld and surrounding stars are rather intermediate in terms of location and age. It therefore seems that star formation in the last $8\text{--}10\,\myr$ started north to north-west of \pulsar and along the molecular filament between NGC~2060 and NGC~2070, and then moved towards the position of the pulsar.

The age distribution of stars in NGC~2060 (Fig.~\ref{fig:age-mass-distr-ngc2060}a) shows that stars are on average older than in NGC~2070, with a median age of $5.7\,\myr$. The dominant feature in the age distribution of stars in NGC~2060, a peak at about $4.4\,\myr$, is also found in our age distribution of all sample stars (Fig.~\ref{fig:age-mass-distr-whole-region}a), but not in that of stars around the R136 star cluster (Fig.~\ref{fig:age-mass-distr-r136}a). The older median age is also represented by, on average, less massive and fewer massive stars: the distribution of initial masses in Fig.~\ref{fig:age-mass-distr-ngc2060}b is truncated above $100\,\msun$ and the fitted power-law exponent is $2.97$, steeper than for stars in NGC~2070. 

The age and mass distributions of stars in NGC~2060 provide probabilistic constraints on the progenitor star that gave rise to the pulsar \pulsar and the supernova remnant N157B. Various studies have attempted to obtain such constraints on other nearby supernova remnants from photometric measurements of the surrounding stellar population; for example see \citet{2014ApJ...791..105W} for a discussion of constraints on the progenitor masses of 17 historical supernovae. The age distribution around N157B is fairly broad, but peaks at ${\approx}\,4.4\,\myr$ (Fig.~\ref{fig:age-mass-distr-ngc2060}a). So, if the progenitor of \pulsar also belonged to stars aged $4\text{--}5\,\myr$, its initial mass is between $35$ and $50\,\msun$, assuming that the star lived its life as a genuine single star.

\subsubsection{Stars outside NGC~2060 and NGC~2070}\label{sec:am-stars-outside}

The age distribution of stars outside NGC~2060 and NGC~2070 (Fig.~\ref{fig:age-mass-distr-outside}a) shows a significant over-abundance of ${\approx}\, 4.4\,\myr$ old stars. A similar over-abundance is found in NGC~2060 (Fig.~\ref{fig:age-mass-distr-ngc2060}), but not in the region around R136 (Fig.~\ref{fig:age-mass-distr-r136}). Also the shoulder at about $6.6\,\myr$ is reminiscent of that in NGC~2060 whereas the $2.5\,\myr$ peak is rather found in R136. The median age of stars in this sample is $8.1\,\myr$, older than the median age of stars in NGC~2060 and NGC~2070, demonstrating that the NGC~2060 and NGC~2070 regions contain on-average the youngest stars in \tdor.

Despite the older median age, the mass distribution (Fig.~\ref{fig:age-mass-distr-outside}b) is filled up to the highest masses (in contrast to the mass distribution of stars in NGC~2060; Fig~\ref{fig:age-mass-distr-ngc2060}b). It is difficult to fit the mass distribution with a single power-law because a flatter slope is required at the high mass end ($\gtrsim 60\,\msun$) than at lower masses. Such a trend is expected because the mass-luminosity exponent $x$ is smaller in more massive stars such that a present-day mass function of a population of stars that underwent constant star formation in the past is flattest at the highest masses (see Sect.~\ref{sec:imf}).

\subsection{Spatial distribution of stellar ages and masses}\label{sec:spatial-distribution}

\begin{figure*}
\begin{centering}
\includegraphics[width=0.95\textwidth]{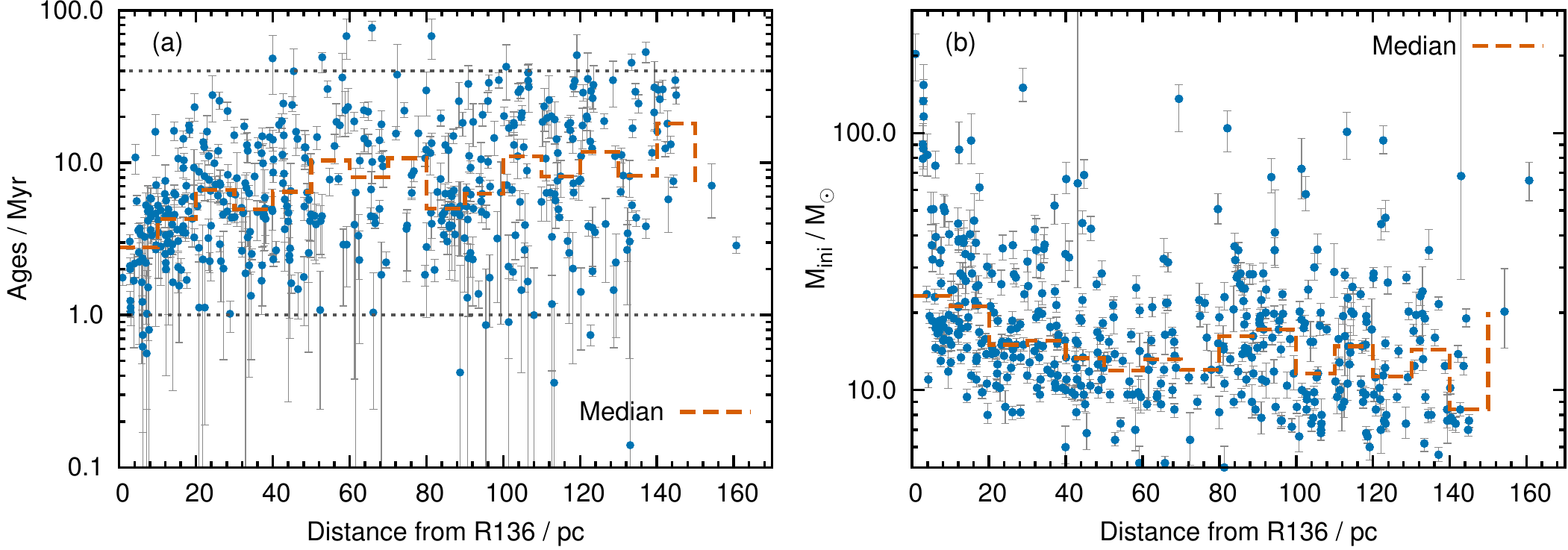}
\par\end{centering}
\caption{Distribution of (a) ages and (b) initial masses of our sample stars as a function of distance from the central R136 star cluster. The dashed lines show the median age and mass in $10\,\mathrm{pc}$ distance intervals and the grey dotted lines highlight ages of $1$ and $40\,\myr$. The median ages and masses have not been corrected for sample completeness (\cf Sect~\ref{sec:completeness-corrections}).}
\label{fig:distances}
\end{figure*}

In Fig.~\ref{fig:distances}, we show the distribution of ages and initial masses of our sample stars as a function of distance from the central R136 star cluster. The age gradient in the close vicinity of R136 is visible, highlighting again that the youngest stars are concentrated towards R136 and a core--halo age gradient is present. In fact, this trend even prevails over the whole \tdor nebula: massive stars are on average older and less massive the further away they are from R136.

Stars of all ages and masses are found across the whole \tdor nebula (\eg Figs.~\ref{fig:age-map-whole-region} and~\ref{fig:distances}). Even stars up to $100\,\msun$ are found far from R136 without clear association to other massive stars in terms of location and age. Stars that were born around the same time in the same region of \tdor would be expected to stand out as `groups' in Fig.~\ref{fig:distances}a and potentially as peaks in our age distributions. At a distance of $80\text{--}100\,\mathrm{pc}$ from R136, there is such a group of stars with an age of $4\text{--}5\,\myr$; the massive stars close to NGC~2060 and north of it (see Fig.~\ref{fig:age-map-ngc2060} and Sect.~\ref{sec:am-ngc2060-region}). The same group of stars has also been identified in our age distributions: the peaks at about $4.4\,\myr$ in the age distributions of stars in NGC~2060 and in the whole \tdor field (Figs.~\ref{fig:age-mass-distr-ngc2060} and~\ref{fig:age-mass-distr-outside}, respectively). In total, there are about 50--60 OB stars in our sample that could belong to this group given their age of ${\approx}\,4.4\,\myr$. However, these stars are scattered around the whole \tdor nebula without an obvious spatial concentration towards NGC~2060, so it seems that not all of them formed in the same location. 

Except for the group of $4.4\,\myr$ old stars, there are no obvious other coeval groups of stars in our sample. The distribution of stellar ages as a function of distance is quite homogeneous and there are no other peaks in our age distributions that are significant at the ${\gtrsim}\,1\sigma$ level. This is not to say that other, approximately coeval groups of stars do not exist in \tdor (\eg \tld and R136), but they are not easily identified with our current data and analysis.

\subsection{Initial mass function of massive stars in \tdor}\label{sec:imf}

Our age and mass distributions of stars in \tdor (Figs.~\ref{fig:age-mass-distr-whole-region}, \ref{fig:age-mass-distr-r136}, \ref{fig:age-mass-distr-ngc2060} and \ref{fig:age-mass-distr-outside}) are not SFHs and IMFs, respectively, because they lack stars that already ended their nuclear burning. Our mass distributions are the convolution of the underlying SFH and IMF. For example, for a constant star-formation rate and a mass-luminosity relation of the form $L\,{\propto}\, M^x$, the slope of the present-day mass-function is $\alpha = \gamma+x-1$, meaning it is steeper than the true IMF slope $\gamma$ because $x\,{\geq}\,1$. The steepening is because of shorter lifetimes associated with more massive stars such that the present-day mass function is depleted from the high mass end onwards. The true IMF slope $\gamma$ of our different stellar samples in \tdor must thus be flatter than the slopes $\alpha$ obtained by fitting the mass distributions with power-law functions.

The SFH and IMF can only be unambiguously derived if both the present-day distributions of ages and initial masses are known. With our data, we are therefore able to infer the SFH and IMF of different samples of VFTS stars. To do so, it is important to consider unbiased samples of stars. As described in \citet{2018Sci...359...69S} and in this work, our sample of massive VFTS stars is unbiased for stars more massive than $15\,\msun$; for lower mass stars, the magnitude limit of the VFTS introduces a bias (see for example the turn-over in the mass function in Fig.~\ref{fig:age-mass-distr-whole-region}b at masses smaller than $15\,\msun$ and the magnitude cut in the HR diagram in Fig.~\ref{fig:hrd}). We follow the approach of \citet{2018Sci...359...69S} to infer the SFH and IMF: we assume a power-law IMF, $\xi(M) \propto M^{-\gamma}$, and vary the IMF slope $\gamma$ until we find the best-fit to the distribution of initial masses as observed for all stars that are presently in our sample and more massive than $15\,\msun$. To that end, we first infer the SFH for a given IMF and then---for the inferred SFH and assumed IMF---predict the distribution of initial masses as observed today which we compare to the observed mass distribution by computing the usual $\chi^2$. From the $\chi^2$ values, we compute probability distributions for the IMF slopes $\gamma$. A summary of the median ages and inferred IMF slopes of the different regions in \tdor are provided in Table~\ref{tab:sample-summaries}.

\begin{table}
\caption{\label{tab:sample-summaries}Total number of stars ($N_\mathrm{tot}$), number of stars more massive than $15\,\msun$ ($N_{\geq 15}$), median age of stars ($\left<t\right>$), inferred IMF slopes of stars more massive than $15\,\msun$ ($\gamma$), and probabilities that the inferred IMF slope is flatter than the Salpeter value of $2.35$ ($P_{\gamma<2.35}$) in various samples of stars in \tdor.}
\centering
\begin{tabular}{lccccc}
\toprule 
Sample & $N_\mathrm{tot}$ & $N_{\geq 15}$ & $\left<t\right>/\myr$ & $\gamma$ & $P_{\gamma<2.35}$\\
\midrule
\midrule 
Full \tdor & 452 & 247 & $5.3$ & $1.90^{+0.37}_{-0.26}$ & 83\% \\
NGC~2070 & 83 & 70 & $3.6$ & $1.65^{+0.63}_{-0.42}$ & 77\% \\
NGC~2060 & 71 & 47 & $5.7$ & $2.00^{+0.72}_{-0.60}$ & 63\% \\
`Field' & 298 & 130 & $8.1$ & $1.90^{+0.69}_{-0.39}$ & 64\% \\
\bottomrule
\end{tabular}\\
\small
`Field' refers to all stars outside NGC~2060 and NGC~2070.
\end{table}

As shown in \citet{2018Sci...359...69S}, the IMF of the VFTS stars is shallower than a Salpeter-like IMF. This raises the question of why this is the case. Most theoretical star-formation models \citep[\eg][]{2002ApJ...576..870P,2005ApJ...630..250K,2008ApJ...684..395H,2012ApJ...761..156F} introduce a characteristic mass for cloud fragmentation and collapse that is closely related to the \citet{1902RSPTA.199....1J} mass, $M_\mathrm{J} \propto \rho^{-1/2} T^{3/2}$. As the density $\rho$ increases, self-gravity becomes more dominant and the resulting stellar mass decreases. Vice versa, when the temperature $T$ increases, such as in regions of strong feedback or when the stabilising effects of small-scale turbulent motions or magnetic fields are included in the analysis, the characteristic stellar mass becomes larger. For a more detailed account, see for example the reviews by \citet{2004RvMP...76..125M}, \citet{2007ARA&A..45..565M} and \citet{2016SAAS...43...85K}. It is conceivable that the energy and momentum input from previous stellar generations in \tdor results in a IMF with a shallower slope than Salpeter \citep[see \eg][]{2005MNRAS.359..211L}. If true, the youngest stellar populations in 30 Dor have formed from the most intensively heated gas and should follow an even shallower IMF. The youngest stars in our sample are found in NGC~2070 around the R136 star cluster and their IMF slope of $\gamma=1.65^{+0.63}_{-0.42}$ appears to be marginally flatter than $\gamma=1.90^{+0.37}_{-0.26}$, which is the value inferred for all stars in 30 Dor (Table~\ref{tab:sample-summaries}). However, in regions of massive star formation, also the density is higher than on average. In addition, at number densities above ${\approx}\,10^5\,\mathrm{cm}^{-3}$ the gas becomes thermally coupled to the dust, which can act as an efficient thermostat due to its high thermal inertia, and so the gas temperature typically does not increase by much \citep{2008ApJ...681..365E}. Consequently, it is often assumed that the competing effects influencing the IMF largely compensate each other, leading to a roughly universal distribution of stellar masses \citep{2005ASSL..329...57E,2011ApJ...731...61E}. Given the large uncertainties in the derived IMF slopes in \tdor, we do not find statistically significant evidence for a spatial dependence of the IMF and thus cannot distinguish between these two scenarios.

We also want to mention the possibility \citep[see also][]{2018Sci...359...69S} that if high-mass stars form via gravitationally focussed mass accretion with mass accretion rates scaling with the square of the mass, that is Bondi--Hoyle--Littleton like accretion \citep{1982NYASA.395..226Z,2001MNRAS.323..785B,2001MNRAS.324..573B}, the power-law index of the mass spectrum of massive stars approaches the asymptotic limit $\gamma\rightarrow2.00$ \citep{1982NYASA.395..226Z,2015MNRAS.452..566B}. This limit can only be reached if stars accrete a substantial fraction of their seed mass which could be the case for the high mass stars considered in this study. In this picture, our inferred IMF slope for high mass stars in \tdor would rather be the norm than the exception.

In an alternative model, it is assumed that the IMF is inherited from the core mass function and that the two are similar \citep[see \eg Section~3.3 in][]{2007ARA&A..45..565M}. \citet{2018NatAs...2..478M} have been able to obtain the core mass function in the Galactic star-formation region W43-MM1 in the mass range $1.6\text{--}100\,\msun$. They find a power-law index of $1.96\,{\pm}\,0.13$ that is similar to our high-mass IMF slope and also flatter than the slope of a Salpeter-like IMF.

\section{Overall star-formation process in \tdor}\label{sec:overall-sf-process}

We start the discussion of the overall star-formation process in \tdor by briefly reviewing the current state of the art (Sect.~\ref{sec:overview-sf-literature}) and considering the influence of binary stars (Sect.~\ref{sec:binary-stars}) and extinction (Sect.~\ref{sec:extinction}) on our work. We have shown above that stars of all ages and masses are scattered all across \tdor (\eg Sect.~\ref{sec:spatial-distribution}). This constitutes an important piece of evidence towards understanding star formation in a giant molecular cloud such as \tdor and raises the question of where stars were born (Sect.~\ref{sec:where-did-stars-form}). We then take all the evidence together and discuss the emerging picture of how massive stars have formed in \tdor (Sect.~\ref{sec:sf-picture-from-vfts}).

\subsection{Overview of star formation in \tdor}\label{sec:overview-sf-literature}

The \tdor region has a complex structure and it remains uncertain what may have triggered the first star formation. It has been suggested that star formation at the leading edge of the LMC disc has been instigated through ram pressure with the Milky Way's hot halo gas \citep{1998A&A...329L..49D}. This may have lead to a general progression of star formation throughout the molecular ridge south from \tdor. In an alternative scenario, it has been suggested that gas from the Small Magellanic Cloud may be accreting onto the LMC disc in the region of \tdor and the molecular ridge, and has thereby triggered star formation \citep{2011ApJ...737...29O,2017PASJ...69L...5F}.

The oldest massive stars in \tdor formed at least $20\text{--}30\,\myr$ ago \citep[\eg][]{1997ApJS..112..457W,2000AJ....119..787G,2011ApJ...739...27D} whereas the youngest stars are still embedded in dense molecular clumps, primarily found to the north and west of R136 \citep[\eg][]{1987ApJ...323L..65W,1992MNRAS.257..391H,1997ApJS..112..457W,1998AJ....116.1708R,2001AJ....122..858B,2002AJ....124.1601W,2005A&A...438..663M,2016ApJ...831...32N,2018ApJ...852...71K}. There are several star clusters such as Hodge~301 \citep[$15\text{--}30\,\myr$, \eg][]{2000AJ....119..787G,2015A&A...574A..13E,2016ApJ...833..154C}, R136 \citep[$1.5^{+0.5}_{-0.7}\,\myr$, \eg][]{2016MNRAS.458..624C}, SL~639 \citep[$10\text{--}15\,\myr$;][]{2015A&A...574A..13E} and \tld (${\approx} 3.5\,\myr$; Sect.~\ref{sec:am-ngc2060-region} and Appendix~\ref{sec:tld1}). \citet{1997ApJS..112..457W} further suggest the existence of a loose association of stars around the luminous blue variable R143, about $2.2\,\mathrm{arcmin}$ ($32\,\mathrm{pc}$) south to south-east of R136, and \citet{2012ApJ...754L..37S} find a second clump of stars just north-est of R136 that they suggest might be in the process of merging with R136.

Using the Atacama Large Millimeter Array and the Australia Telescope Compact Array, \citet{2013ApJ...774...73I}, \citet{2014ApJ...793...37A} and \citet{2016ApJ...831...32N} find molecular clumps in the filamentary structure around R136 that seem to be gravitationally unstable and are likely to collapse to form the next generation of stars. This is in the same region where \citet{1987ApJ...323L..65W} already identified knots of embedded young stars and where \citet{2018ApJ...852...71K} suggest that stars currently form. Also maser emission is found in these regions, indicative of young massive objects hidden from direct view \citep[\eg][]{2001ApJ...547L..61V,2006MNRAS.372.1509O}.

Pre-MS stars trace recent star formation and are found all over \tdor with clear over-densities in and around NGC~2060 and NGC~2070 \citep[\eg][]{2016ApJS..222...11S,2018MNRAS.tmp.1265K}. \citet{2011ApJ...739...27D} find that pre-MS stars younger than $4\,\myr$ primarily cluster around R136 and towards its north whereas those older than about $12\,\myr$ are rather found east and south-east of R136. These two groups overlap in the outskirts of R136, highlighting the 3D structure of the \tdor nebula.

From photometry of stars near R136, \citet{1996ApJ...466..254B} and \citet{1999A&A...347..532S} determine stellar age distributions similar to ours for stars in NGC~2070. They also find that the region around R136 consists of several stellar populations. \citet{1999A&A...347..532S} find that the youngest stars are concentrated towards the core of R136, in agreement with our results. Similar core--halo age gradients are found in other star forming regions, for example in the Orion nebula, the Flame nebula (NGC~2024) and W40 \citep{2014ApJ...787..109G,2014ApJ...787..108G}. They even occur in 80\% of young star clusters studied by \citet{2018MNRAS.tmp..296G} and thus appear to be a ubiquitous feature of star formation.

\subsection{Influence of binary stars}\label{sec:binary-stars}

Binary stars influence our work in two ways: (i) unrecognised binaries can lead to overestimated luminosities and hence masses \citep[see discussion in][]{2018Sci...359...69S}, and (ii) mass can be exchanged in binaries (\eg stellar mergers), which increases the masses of stars and leads to rejuvenation, that is to underestimated ages \citep[\eg][]{1983Ap&SS..96...37H,1992ApJ...391..246P,2012ARA&A..50..107L,2016MNRAS.457.2355S}. These two aspects need to be kept in mind when interpreting our data and drawing conclusions from it.

\citet{2018Sci...359...69S} show that both unrecognised binaries and binary mass-exchange products have a negligible effect on the inferred IMF slopes in our case, because the VFTS is a multi-epoch spectroscopic survey where the binary detection probability increases with mass, shorter orbits and larger mass ratios. This means that those binaries that would affect the inference of the IMF most strongly are more likely to be detected and hence removed from our sample. 

However, the ages and masses of individual stars, and also the raw age and mass distributions derived in this work are affected by binary stars. \citet{2014ApJ...782....7D} estimate that about 30\% of stars in a sample similar to ours are products of binary interactions. This implies that the ages of about one third of our sample stars are potentially underestimated because of rejuvenation. This needs to be accounted for when drawing conclusions about the star formation process in \tdor.

\subsection{Extinction}\label{sec:extinction}

The visual extinction towards OB stars in \tdor is relatively low, though is inhomogeneous across the region \citep{2014A&A...564A..63M}. \citet{2016MNRAS.455.4373D} have quantified the spread in visual extinction across \tdor ($A_{555}$ of $0.5$ to $3.0\,\mathrm{mag}$) which needs to be kept in mind when discussing the average ages of our OB stars in different regions because of potentially varying completeness of the VFTS with location in \tdor. For example, \citeauthor{2016MNRAS.455.4373D} find that there is more extinction towards the south-east of NGC~2060 and NGC~2070 than towards the north-west.

The mass distributions of our sample stars in NGC~2070 (Fig.~\ref{fig:age-mass-distr-r136}b), NGC~2060 (Fig.~\ref{fig:age-mass-distr-ngc2060}b) and outside these regions (Fig.~\ref{fig:age-mass-distr-outside}b) show the same transition from a plateau to a power-law function at about $15\,\msun$ such that it seems that differences in extinction between these regions are on average not so important for stars ${\gtrsim}\,15\,\msun$. However, if the average extinction is indeed larger in NGC~2060 than in NGC~2070, we might lack more BA stars in NGC~2060 than in NGC~2070 because of the magnitude limit of the VFTS. If true, this would make the median age of NGC~2060 older, reinforcing our finding that stars in NGC~2060 are on average older than those in NGC~2070.

\subsection{Where did massive stars form?}\label{sec:where-did-stars-form}

There are clear overdensities of massive stars in NGC~2060 and NGC~2070, indicating that the bulk of these stars formed in these two regions. But where did the stars outside of NGC~2060 and NGC~2070 form? They could have formed close to where they are observed today or travelled to their current positions from NGC~2060 and NGC~2070. To test the relevance of these formation channels in \tdor (\ie star formation in relative isolation vs.\ star formation in clusters and associations), we consider three probes: (i) similarities between our age distributions of different regions in \tdor, (ii) the radial-velocities of the VFTS stars and (iii) the proper motions of some VFTS stars from the second \gaia data release (\gaia DR2).

If stars were to move fast enough such that they are strongly mixed, the age distributions derived for sub-regions in \tdor would show similar features. From our age distributions of the different studied sub-regions (Figs.~\ref{fig:age-mass-distr-whole-region}a, \ref{fig:age-mass-distr-r136}a, \ref{fig:age-mass-distr-ngc2060}a and~\ref{fig:age-mass-distr-outside}a), it is evident that this is not necessarily the case. There are features in the age distributions that appear to be unique to certain sub-regions, such as the most dominant feature at ${\approx}\,3.5\,\myr$ in the age distribution of stars in NGC~2070 (Fig.~\ref{fig:age-mass-distr-r136}a), which neither stands out in the age distribution of stars in NGC~2060 (Fig.~\ref{fig:age-mass-distr-ngc2060}a) nor in that of stars outside NGC~2060 and NGC~2070 (Fig.~\ref{fig:age-mass-distr-outside}a). The same holds true for the group of $4.4\,\myr$ stars which are rather found outside of NGC~2070 and not within it. Also, the age distributions of stars in NGC~2060 and NGC~2070 are qualitatively and quantitatively different, suggesting that the bulk of stars in NGC~2060 and NGC~2070 formed locally and that only a limited amount of mixing occurred between these two regions.

The $1\sigma$ radial-velocity (RV) dispersion of VFTS O-type stars in NGC~2060 and NGC~2070 is ${\approx}\,8.0\,\kms$ and ${\approx}\,8.6\,\kms$, respectively \citep{Sana+2014}, and our sample contains 27 RV runaway candidates, that is stars with a radial velocity of ${\gtrsim}\,25.8\,\kms$ (indicated by crosses in Figs.~\ref{fig:age-map-whole-region}, \ref{fig:age-map-R136} and~\ref{fig:age-map-ngc2060}). With such velocities, stars can travel considerable distances within typical lifetimes of a few Myr and thus contribute to the dilution and mixing of stellar populations in \tdor. 

\begin{figure}
\begin{centering}
\includegraphics[width=0.48\textwidth]{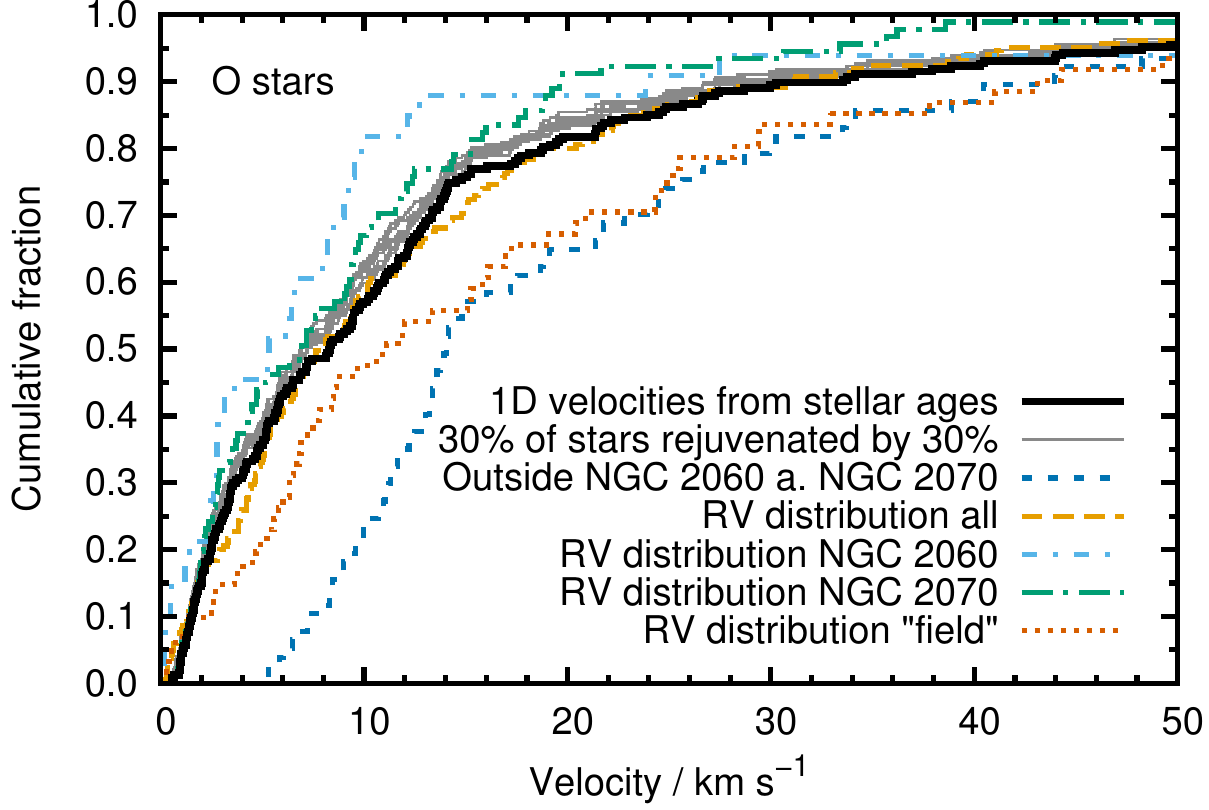}
\par\end{centering}
\caption{Cumulative distribution of O-star velocities ($M_\mathrm{ini}\,{\geq}\,15\,\msun$) in \tdor. The comparison radial-velocity (RV) distributions are from \citet{2013A&A...550A.107S}. The term `field' refers to stars outside NGC~2060 and NGC~2070, and we show ten realisations of the inferred velocity distribution under the assumption that 30\% of stars are rejuvenated binary products that appear younger by 30\% compared to their true age (solid thin grey lines).}
\label{fig:vel-distr}
\end{figure}

To further quantify how far stars might have migrated in \tdor, we compute the transverse velocities of stars required to move to their current position from R136 given the inferred stellar ages. It is important to account for rejuvenation (Sect.~\ref{sec:binary-stars}) and we assume that our sample contains 30\% of rejuvenated binary products \citep{2014ApJ...782....7D} which appear to be 30\% younger than they truly are \citep[this is a characteristic average of rejuvenation in stellar mergers;][]{2016MNRAS.457.2355S}. To avoid biases, we focus on stars with an initial mass ${\geq}\, 15\,\msun$. For comparison with the distribution of radial velocities of VFTS single O stars from \citet{2013A&A...550A.107S}, we convert the 2D transverse velocities into their 1D equivalent by dividing by $\sqrt{2}$. The resulting cumulative velocity distributions of all stars and those that are $2.4\,\mathrm{arcmin}$ outside NGC~2060 and NGC~2070 are shown in Fig.~\ref{fig:vel-distr} alongside the RV distributions for stars in various sub-regions of \tdor. We have chosen a slightly different definition of NGC~2060 and NGC~2070 than in the rest of this work for consistency with \citet{Sana+2014}; this does not affect our conclusions.

To reach their current positions from R136 within the inferred ages, our sample stars require faster velocities than expected by the RV distributions of stars in NGC~2060 and NGC~2070, even when accounting for rejuvenation. This suggests that some stars outside NGC~2060 and NGC~2070 formed in situ. However, there is also the possibility that the RV distributions of stars in NGC~2060 and NGC~2070 are biased against fast velocities because such stars could have potentially left these regions already. It is therefore instructive to also compare our inferred velocities to RVs of stars outside these two regions. Indeed, the RV distribution is shifted to higher velocities (Fig.~\ref{fig:vel-distr}), indicative of a larger contribution of fast moving stars outside NGC~2060 and NGC~2070. While the RV distribution of stars with velocities ${\gtrsim}\, 14\,\kms$ appears broadly consistent with the distribution of velocities of stars outside NGC~2060 and NGC~2070 inferred from the stellar ages, the two distributions differ at lower velocities. There are O stars outside NGC~2060 and NGC~2070 whose RV distribution predicts too slow velocities to reach their observed positions in \tdor within the inferred ages if these stars formed in R136. This again indicates that some stars most likely formed in situ outside NGC~2060 and NGC~2070. 

In general, the RV distribution of VFTS O stars in \tdor agrees with the inferred velocity distribution from the stellar ages. This demonstrates that O stars in \tdor can travel over large distances and that this contributes to the overall mixing of stellar populations.

\begin{figure*}
\begin{centering}
\includegraphics[width=0.95\textwidth]{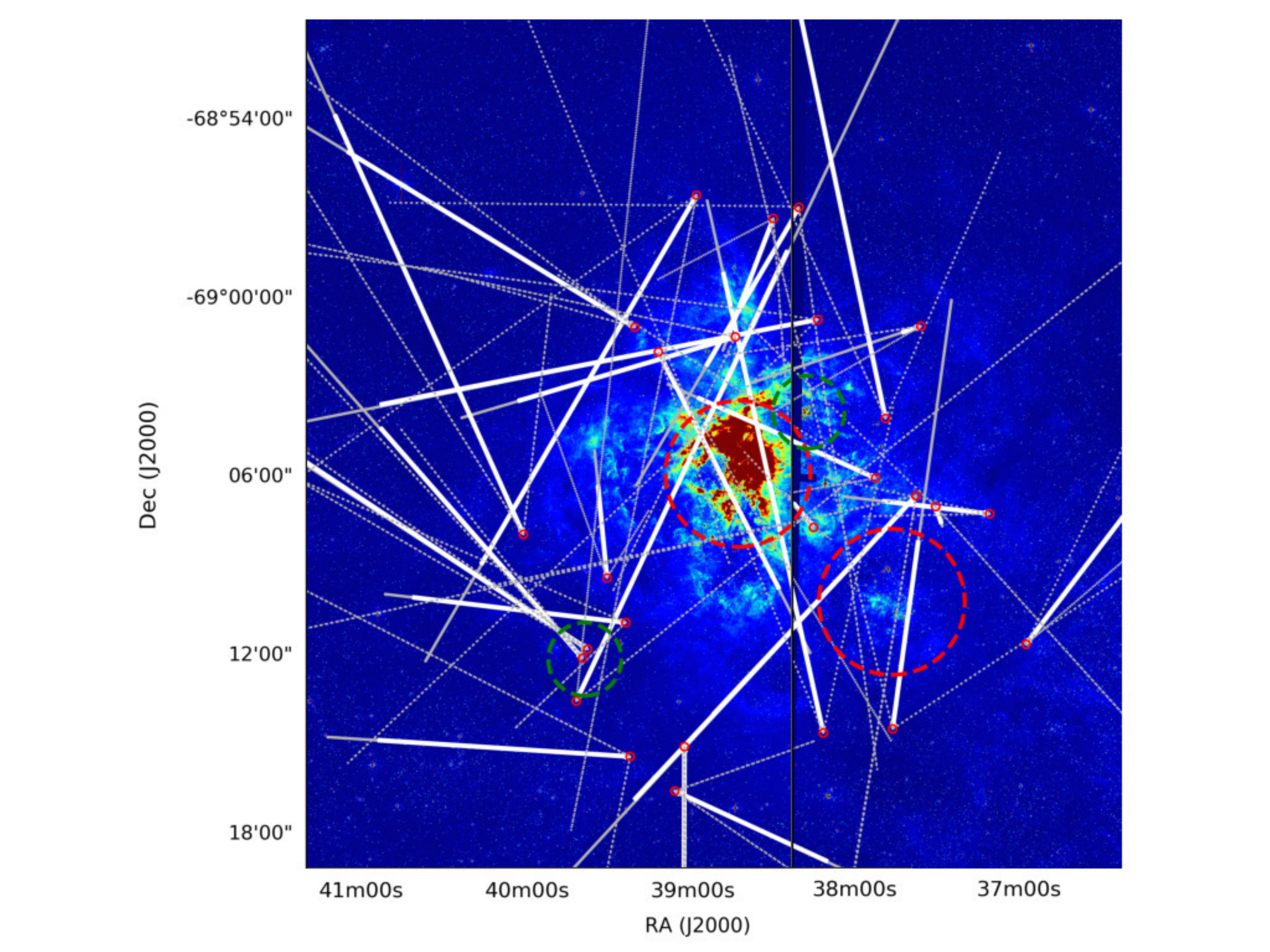}
\par\end{centering}
\caption{Backtracing the birth places of some VFTS O stars outside NGC~2060 and NGC~2070 with \gaia DR2 proper motions. The thick lines indicate the past movement of stars given their inferred stellar ages and the `cones' (dotted lines) represent the range of possible motions because of proper-motion uncertainties. The longer thin lines represent the larger travelled distances from the age uncertainties. The red circles show NGC~2060 and NGC~2070 while the green circles show Hodge~301 and SL~639. The background image is a ESO WFI composite based on observations made with ESO Telescopes at the La Silla Observatory under programme ID 076.C-0888, processed and released by the ESO VOS/ADP group.}
\label{fig:gaia-map}
\end{figure*}

Thanks to the \gaia satellite and its second data release \citep{2016A&A...595A...1G,2018arXiv180409365G}, there is information available on the proper motions of stars in \tdor. We cross-correlate \gaia targets with VFTS stars more massive than $15\,\msun$ for which we have determined stellar ages. We exclude RV runaway candidates, exclude stars closer than $3\,\mathrm{arcmin}$ to NGC~2060 and NGC~2070, and select those \gaia targets for which proper motions are known to better than $0.07\,\mathrm{mas}\,\mathrm{yr}^{-1}$ in right-ascension and declination (corresponding to ${\lesssim}\,20\,\kms$ at a distance of $50\,\mathrm{kpc}$ to the LMC; the chosen proper-motion threshold is arbitrary and does not affect the conclusions). In Fig.~\ref{fig:gaia-map}, we show the positions of these stars and a `cone' indicating their likely birth places computed from the proper motions, their uncertainties and the inferred ages. Already from this rather limited sample of stars, it is evident that there are several stars in our sample whose past movements cannot be traced back to NGC~2060 and NGC~2070, implying that they formed outside these two regions. Some \gaia proper motions of stars in \tdor in the second data release are probably not yet final and might change in the next data release. In particular, fully understanding the correlations in the proper motions is challenging. In a forthcoming paper, this aspect will be investigated in more detail and some of the proper motions in Fig.~\ref{fig:gaia-map} might need to be updated. However, we do not expect that this will affect our general conclusions from the current data.

The idea that some massive stars form across the whole \tdor field is further supported by \citet{2012A&A...542A..49B}, who find several candidates in the VFTS sample that fulfil their criteria for star formation in apparent isolation, and by \citet{2014A&A...564L...7K}, who find a massive (${\approx}\,11\,\msun$) pre-MS candidate in the outskirts of \tdor. Also, pre-MS stars are found all over \tdor and not only in NGC~2060 and NGC~2070 \citep[\eg][]{2018MNRAS.tmp.1265K}, suggesting that stars formed across the whole \tdor nebula and not exclusively in dense associations and clusters. This is also in agreement with the work of \citet{2018MNRAS.475.5659W}, who find that large fractions of stars in Galactic OB associations appear to rather form in situ than in dense cluster-like environments. A similar conclusion has been reached by \citet{2016MNRAS.460.2593W} for stars in the massive Cygnus~OB2 association. We therefore conclude that the massive star population outside NGC~2060 and NGC~2070 consists of two components: runaways and walkaways, and stars that formed in situ.

\subsection{Picture of massive star formation from the VFTS}\label{sec:sf-picture-from-vfts}

The oldest stars in our sample have ages of ${\gtrsim}\,50\,\myr$ (Fig.~\ref{fig:distances}a). With a velocity of $10\,\kms$, such stars could have travelled a distance of about $500\,\mathrm{pc}$, so it is not clear whether these stars formed in \tdor or are interlopers from nearby star-forming regions. Our sample of stars is only unbiased for stars more massive than $15\,\msun$. Such stars have a nuclear burning lifetime of about $12\,\myr$, which sets the oldest age up to which we can constrain the SFH. From deeper photometric observations, \citet{2015ApJ...811...76C} conclude that the star-formation rate in \tdor more than ${\approx}\,20\,\myr$ ago was indistinguishable from the average star formation in the LMC and has increased since then. This coincides with our results that the number of stars in our sample starts to increase at ages of ${\lesssim}\,40\,\myr$ and has reached a considerable level at ages of ${\lesssim}\,20\,\myr$ (Fig.~\ref{fig:distances}a). The star-formation rate then increased rapidly $8\,\myr$ ago as evident from our age distribution (Fig.~\ref{fig:age-mass-distr-whole-region}a) and the SFH of massive VFTS stars derived in \citet{2018Sci...359...69S}.

There is a lack of stars younger than $1\,\myr$ in our sample (Fig.~\ref{fig:distances}a) and also stars in the R136 star cluster are most likely older than ${\approx}\,1\,\myr$ (Sect.~\ref{sec:am-r136-region}). This suggests that either stars younger than $1\,\myr$ are still hidden from our view (\eg they might still be embedded in their birth clouds; see also Sect.~\ref{sec:extinction}) or that star formation ceased with the birth of R136. At least there is currently no evidence for a second R136-like star cluster hidden in \tdor \citep{2016ApJ...821...51R} such that it seems reasonable to assume that the overall star-formation rate has dropped after giving birth to the bulk of very massive stars in R136 about $1\text{--}2\,\myr$ ago. One could speculate whether the birth of very massive stars in R136 might be responsible for suppressing star formation through their strong feedback locally in NGC~2070, but this can of course not explain the apparent lack of stars younger than $1\,\myr$ throughout \tdor (\cf Fig.~\ref{fig:distances}a). Unfortunately, also \citet{2015ApJ...811...76C} are currently unable to probe the SFH in the last ${\approx}\,1\,\myr$ from their deeper observations of lower mass stars in \tdor.

Between $1.5$ and $5\,\myr$, our age distribution of the full \tdor region shows a plateau (Fig.~\ref{fig:age-mass-distr-whole-region}). This is in agreement with the star-formation history of considerably lower mass stars in NGC~2070 as inferred by \citet{2015ApJ...811...76C}. The formation history of low and high mass stars therefore appears to be similar and no obvious delay between their formation is found. 
 
Taking all our evidence together, the following picture of star formation in \tdor emerges. First, stars formed in the \tdor field outside NGC~2060 and NGC~2070. We observe an age gradient in NGC~2060 (Sect.~\ref{sec:am-ngc2060-region}) that implies that star formation started in the north and north-west of NGC~2060 and then proceeded inwards towards the position of the pulsar \pulsar. More or less at the same time, star formation also proceeded into the NGC~2070 region, but lasted longer than in NGC~2060 (giving rise to the younger median age) and culminated in the formation of the R136 star cluster. In this scenario, R136 formed last and thereby produced the core--halo age structure in NGC~2070. A considerable fraction of stars have been ejected and/or moved away from their birth places and thereby contribute to the formation of the age gradient across \tdor (stars being more spread out the older they are) and to the large age ranges found all over \tdor. Residual star formation still takes place in the north and north-west of NGC~2070 and some of it may have even been triggered by the energetic feedback of young stars in R136 as suggested for example by \citet{1997ApJS..112..457W}, but we cannot test this hypothesis with our data.

\begin{figure}
\begin{centering}
\includegraphics[width=0.48\textwidth]{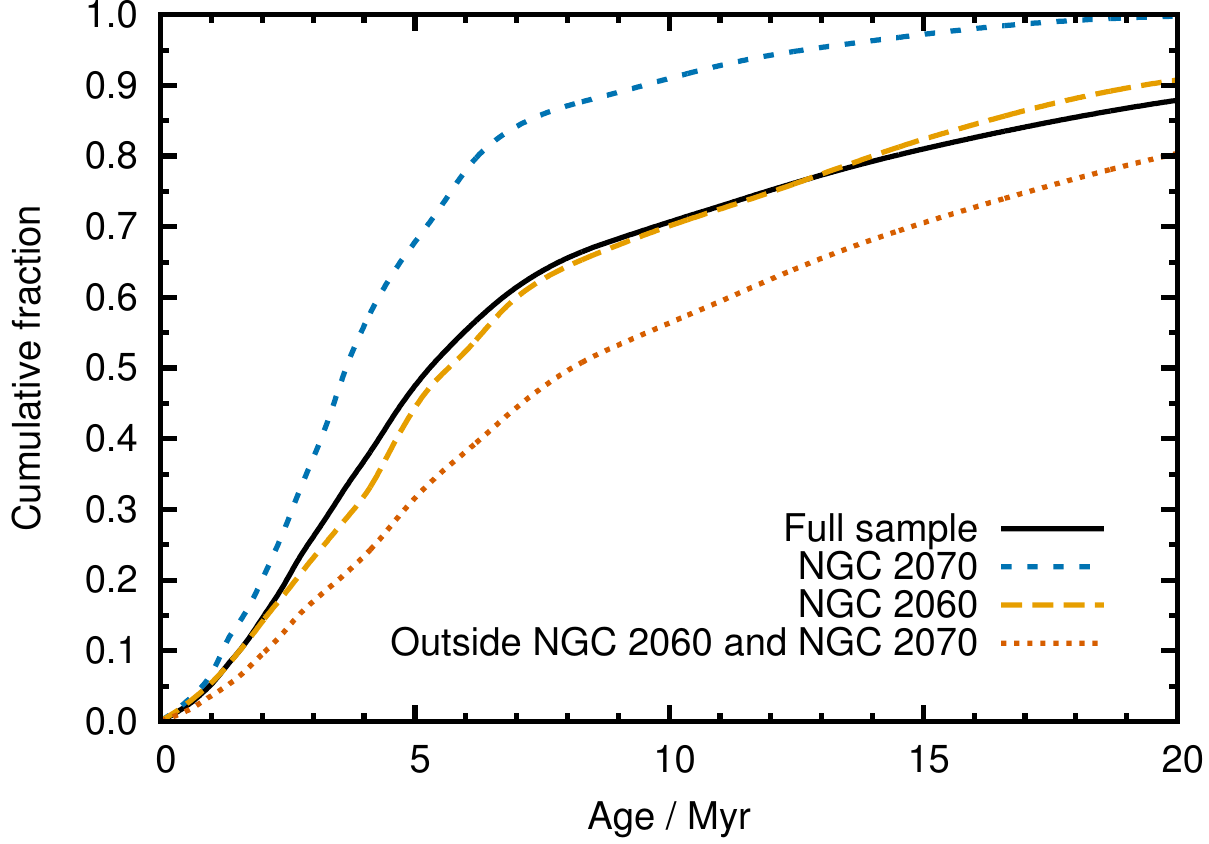}
\par\end{centering}
\caption{Cumulative age distributions of stars in our full sample, in NGC~2070, in NGC~2060 and outside these two regions. The different average ages of these samples are evident (\cf Table~\ref{tab:sample-summaries}) and also the ages at which the star formation rate increased rapidly (kinks in the distributions).}
\label{fig:cumulative-age-distr}
\end{figure}

This sequence of star formation is evident from the median ages of stars in NGC~2070, in NGC~2060 and outside these two regions (Table~\ref{tab:sample-summaries}), and even more so from the cumulative age distributions of stars in these regions of \tdor (Fig.~\ref{fig:cumulative-age-distr}). The kinks in the cumulative age distributions indicate the onset of the rapid increase in star formation and thereby trace the sequence of how stars formed in \tdor. It appears that the rapid increase of star formation is offset by ${\approx}\,2\,\myr$ in the different regions of \tdor (see also Figs.~\ref{fig:age-map-whole-region}a, \ref{fig:age-mass-distr-r136}a, \ref{fig:age-mass-distr-ngc2060}a and~\ref{fig:age-mass-distr-outside}a), but is overall quite synchronised.

Simulations of the collapse of (giant) molecular clouds show fragmentation and the formation of filaments, sheets and clumps where stars subsequently form \citep[\eg][]{1978MNRAS.184...69L,1998ApJ...501L.205K,2000ApJS..128..287K,2003MNRAS.343..413B,2003MNRAS.339..577B,2009MNRAS.392..590B,2011ApJ...740...74K,2012ApJ...754...71K,2012MNRAS.424.2599C,2012MNRAS.424..377D,2013MNRAS.430..234D,2014prpl.conf..291L,2017MNRAS.467.1313V}. Following this idea, we envisage that the \tdor giant molecular cloud began to fragment and form stars in a relatively unordered way over the whole \tdor field. The biggest concentrations of stars would form where filaments and sheets intersect and thus produce the biggest gravitational troughs into which gas and stars might be further channelled. Depending on the exact conditions, clusters and associations of stars can form in such places (\eg NGC~2070 including R136, Hodge~301, SL~639 and \tld). The growth of such structures likely takes some time after star formation starts to accelerate in a giant molecular cloud. One might therefore expect to find time delays between the formation of these structures. This could be an explanation for the slightly different onsets of the rapid increase of star formation in NGC~2070, NGC~2060 and outside these regions (Fig.~\ref{fig:cumulative-age-distr}).

The biggest star formation event would take place in the region where there is the most dense gas available. Once such a structure forms, it may collapse hierarchically and accrete further gas and stars from farther outside, thereby continuing its growth \citep[\cf][]{2003MNRAS.343..413B,2009MNRAS.400.1775S,2017MNRAS.467.1313V}. NGC~2070 with the R136 star cluster may have formed in this way, continuously producing stars in partly sub-structured clumps and groups (\eg the north-east clump) that may even merge with each other. This could explain the large age range of massive stars found even in this part of \tdor. Because of the continuous supply of gas from outside regions, for example via filaments, the most massive stars could form in the deepest part of the gravitational potential, that is in the core of the R136 star cluster. As suggested by \citet{2018MNRAS.473L..11R}, the birth of this youngest population of very massive stars may have partly terminated the gas accretion and push material out to form the shell-like structure that nowadays surrounds R136. A core--halo age gradient may form naturally in this way because the latest episode of star formation would have occurred in the innermost regions and the oldest stars would have had the most time to disperse \citep{2017MNRAS.467.1313V,2018MNRAS.tmp..296G}. 

A prediction of this formation mode of NGC~2070 and the central R136 star cluster is that even the core of R136 may host stars of different ages because it could have formed from either merging sub-structures or accreting older stars alongside gas from further outside. There indeed appear to be a few apparently old stars in the R136 core \citep{2016MNRAS.458..624C}, but more robust stellar parameters of these stars are required to properly probe this hypothesis. Another piece of evidence for the formation of the innermost NGC~2070 region via mergers of sub-clusters and groups of stars may be the inferred ordered rotation of stars within $10\,\mathrm{pc}$ of R136 \citep{2012A&A...545L...1H} and the north-east clump close to R136 \citep{2012ApJ...754L..37S}.

\section{Summary and conclusions}\label{sec:conclusions}

We study the massive star content and the star formation process in the \tdor nebula in the LMC utilising a sample of 452 (mainly) OB stars from the VLT-FLAMES Tarantula Survey. Stellar parameters have been determined by modelling the observed VFTS spectra and matching the inferred spectroscopic parameters against stellar models using the Bayesian code \bonnsai. The inferred full posterior probability distributions of stellar ages and initial masses for each star in our sample are then combined to investigate age distributions and mass functions of sub samples of massive stars. This unprecedented sample of stars offers the unique possibility to study the evolution of (very) massive stars and the star formation process in the local \tdor starburst region over the last ${\approx}\,10\,\myr$. Our main conclusions can be summarised as follows:
\begin{itemize}
\item The single star models of \citet{2011A&A...530A.115B} and \citet{2015A&A...573A..71K} likely lack physics to explain the positions of very massive stars ($\log L/\lsun\, {\gtrsim}\, 6.0$) in the HR diagram and their helium enriched surfaces \citep[see also][]{2014A&A...570A..38B}. This should hold true for basically all current massive star models \citep[\eg][]{2012A&A...537A.146E,2015MNRAS.452.1068C,2016ApJ...823..102C}. Higher mass-loss rates for stars approaching the Eddington limit might offer a solution to lift the current tension. Also, a Hertzsprung--gap is visible that suggests that the empirical TAMS is at slightly cooler temperatures than that predicted by the above mentioned models. This nicely illustrates the power of the VFTS that allows us for the first time to observationally constrain, improve and calibrate the physics of such very massive star models.
\item There are no obvious, spatially-coherent age-patterns of massive stars in \tdor. This is remarkable because it implies that stars migrated over long distances and/or formed throughout the \tdor nebula in relative isolation. 
\item We find that some massive stars indeed formed in relative isolation across the whole \tdor field and not exclusively in NGC~2060 and NGC~2070, that is not in dense clusters and associations.
\item Massive star formation accelerated about $8\,\myr$ ago, first forming stars in the wider \tdor field (median age $8.1\,\myr$) and later also inside NGC~2060 (median age $5.7\,\myr$) and NGC~2070 (median age $3.6\,\myr$). The R136 star cluster in NGC~2070 stands at the end of this formation process and (residual) star formation is still ongoing in dense molecular knots towards its north and west.
\item Stars in the central starburst region NGC~2070 formed over several $\myr$, with the R136 star cluster forming last and being surrounded by, on average, older stars such that a core--halo age gradient is visible. Such structures are also seen in other star forming regions and may be a ubiquitous feature of star cluster formation.
\item The stellar mass functions of our sample stars are well sampled up to $200\,\msun$. The inferred IMF of stars ${>}\,15\,\msun$ is shallower than a Salpeter-like IMF, as demonstrated by \citet{2018Sci...359...69S}. Within our quite large IMF slope uncertainties, we cannot find a significant spatial dependence of the IMF. However, our data may suggest that the IMF is flattest in NGC~2070, that is around the R136 star cluster.
\end{itemize}

The present study extends and complements previous work \citep[\eg][]{1999A&A...347..532S,2001AJ....122..858B,2011ApJ...739...27D,2015ApJ...811...76C,2016ApJS..222...11S} on the formation of stars in \tdor. Taking all efforts together, a complex and hierarchical picture of star formation emerges where stars form both in relative isolation and in dense star clusters throughout a giant molecular cloud over a relatively long period of time. Our work will greatly help understand the complicated formation process of massive stars and star clusters, and sets a new benchmark for theoretical studies of star formation in giant molecular clouds.

\begin{acknowledgements}
While writing this paper, we received the sad news that our dear friend Dr Nolan Walborn passed away. We are sure he would have loved to see the final version of this paper that discusses a topic that was always very close to his heart. 
We thank the referee for valuable comments that helped improve the paper.
This work was supported by the Oxford Hintze Centre for Astrophysical Surveys which is funded through generous support from the Hintze Family Charitable Foundation.
OHRA acknowledges funding from the European Union's Horizon 2020 research and innovation programme under the Marie Sk{\l}odowska-Curie grant agreement No 665593 awarded to the Science and Technology Facilities Council.
HS acknowledges support from the FWO-Odysseus programme under project G0F8H6N. 
CS-S acknowledges support from CONICYT-Chile through the FONDECYT Postdoctoral Project No.~3170778. 
SSD and AH thank the Spanish MINECO for grants AYA2015-68012-C2-1 and SEV2015-0548, and are supported by Gobierno de Canarias under project ProID2017010115.
GG acknowledges financial support from the Deutsche Forschungsgemeinschaft, Grant No.\ GR 1717/5. 
SdM has received funding under the European Union’s Horizon 2020 research and innovation programme from the European Research Council (ERC, Grant agreement No.~715063)
RGI thanks the STFC for funding his Rutherford fellowship under grant ST/L003910/1 and Churchill College, Cambridge for his fellowship and access to their library.
RSK acknowledges support from DFG in SFB 881 `The Milky Way System' (subprojects B1, B2, and B8) and SPP 1573 `Physics of the ISM' (grant numbers KL 1358/18.1, KL 1358/19.2), he also receives funding from the ERC via advanced grant 339177.
NM acknowledges the financial support of the Bulgarian NSF under grant DN08/1/13.12.2016.
STScI is operated by AURA, Inc. under NASA contract NAS5-26555. 
The raw VFTS observations are available from the European Southern Observatory's Science Archive Facility at \href{http://archive.eso.org}{http://archive.eso.org} under project ID 182.D-0222.
This work has made use of data from the European Space Agency (ESA) mission \gaia (\href{https://www.cosmos.esa.int/gaia}{https://www.cosmos.esa.int/gaia}), processed by the \gaia Data Processing and Analysis Consortium (DPAC, \href{https://www.cosmos.esa.int/web/gaia/dpac/consortium}{https://www.cosmos.esa.int/web/gaia/dpac/consortium}). Funding for the DPAC has been provided by national institutions, in particular the institutions participating in the \gaia Multilateral Agreement.
\end{acknowledgements}

\bibliographystyle{aa}

\appendix

\section{Mass discrepancy}\label{sec:mass-discrepancy}

\citet{1992A&A...261..209H} noted a discrepancy in evolutionary and spectroscopic masses of O-type stars in the sense that evolutionary masses are on average higher than spectroscopic ones. Evolutionary masses are usually derived from a comparison of stars to evolutionary models in the HR diagram, and spectroscopic masses, $M_\mathrm{spec}=1/(4\pi\sigma G)\, (gL/T_\mathrm{eff}^4)$, solely from surface gravities $g$, luminosities $L$ and effective temperatures $T_\mathrm{eff}$ determined from modelling observed spectra. This mass discrepancy has since then been discussed for various samples of stars with different outcomes regarding its existence and magnitude \citep[\eg][]{2002A&A...396..949H,2005ApJ...627..477M,2010A&A...524A..98W,2015A&A...577A..23M,2015IAUS..307..117M,2018A&A...613A..12M}. It has also been discussed within the series of VFTS papers for different stellar samples and we refer the reader to these papers for more details \citep{2015A&A...575A..70M,2017A&A...600A..81R,2017A&A...601A..79S}.

In principle, evolutionary and spectroscopic masses should agree. Comparing the two should therefore rather be a consistency check for the methods and models used to derive spectroscopic and evolutionary masses. There is one situation in which a mass discrepancy is theoretically expected: main-sequence stars with an unusually high helium content will be more luminous than stars of the same mass with a normal helium content. Hence, deriving evolutionary masses for such unusually bright stars with helium-normal stellar models gives too large masses. Such a situation may be encountered when deriving evolutionary masses for helium-enriched binary products with single star models \citep[see \eg][]{2014A&A...564A..52L}.

Other reasons for mass discrepancies must lie within the analysis techniques and applied models, and must be of systematic nature because they would otherwise average out when investigating larger samples of stars. Some potential issues are systematic offsets in the distance to stars, differential extinction, biases in surface gravities derived from atmosphere codes (\eg because of neglected turbulent pressure), convective core-overshooting parameters of stellar models not suited for the observed stars and also the methodology with which evolutionary masses are derived (by-eye comparison of stars in a HR diagram, statistical analysis, choice of observables and prior distributions etc.); a more detailed discussion can for example be found in \citet{2018A&A...613A..12M}. Convective core overshooting most likely has only a very limited effect on the mass discrepancy (N.~Grin private communication).

\begin{figure}
\begin{centering}
\includegraphics[width=0.48\textwidth]{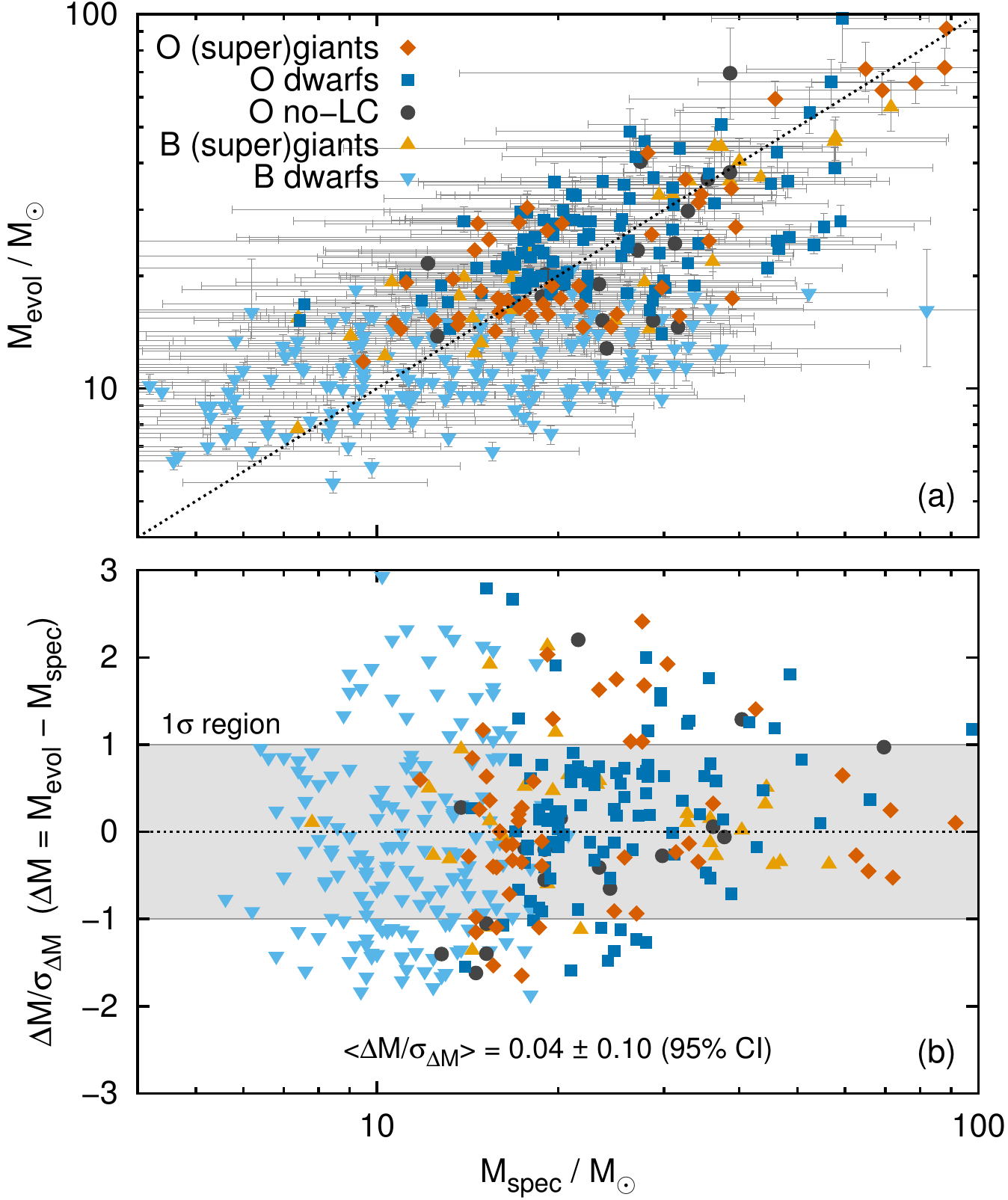}
\par\end{centering}
\caption{Comparison of spectroscopic, $M_\mathrm{spec}$, and evolutionary masses, $M_\mathrm{evol}$, for stars in our sample for which surface gravity, luminosity and effective temperature are known; $\Delta M=M_\mathrm{evol}-M_\mathrm{spec}$ is the mass difference and $\sigma_{\Delta M}$ is the $1\sigma$ uncertainty of $\Delta M$. `O no-LC' refers to stars in our sample with an O-type spectral type but so far unknown luminosity class.}
\label{fig:mass-discrepancy}
\end{figure}

Surface gravity, luminosity and effective temperatures are used for 376 stars with spectroscopic masses larger than $4\,\msun$ (we use a $4\,\msun$ mass-cut to avoid biases because our applied stellar models do not extend to lower masses) to obtain their ages and evolutionary masses. In Fig.~\ref{fig:mass-discrepancy}a we compare spectroscopic ($M_\mathrm{spec}$) and evolutionary ($M_\mathrm{evol}$) masses, and in Fig.~\ref{fig:mass-discrepancy}b we show the normalised mass discrepancy $\Delta M/\sigma_{\Delta M}=(M_\mathrm{evol}-M_\mathrm{spec})/\sigma_{\Delta M}$ as a function of spectroscopic mass (here, $\sigma_{\Delta M}$ is the $1\sigma$ uncertainty of the mass difference $\Delta M$). Overall we find $\Delta M/\sigma_{\Delta M}=0.04{\pm}0.10$ with $0.10$ being the 95\% confidence interval of the standard error of the mean (not to be confused with the dispersion in the mass discrepancy of the sample), meaning that we do not have evidence for a statistically significant mass discrepancy. 

\begin{figure*}
\begin{centering}
\includegraphics[width=0.95\textwidth]{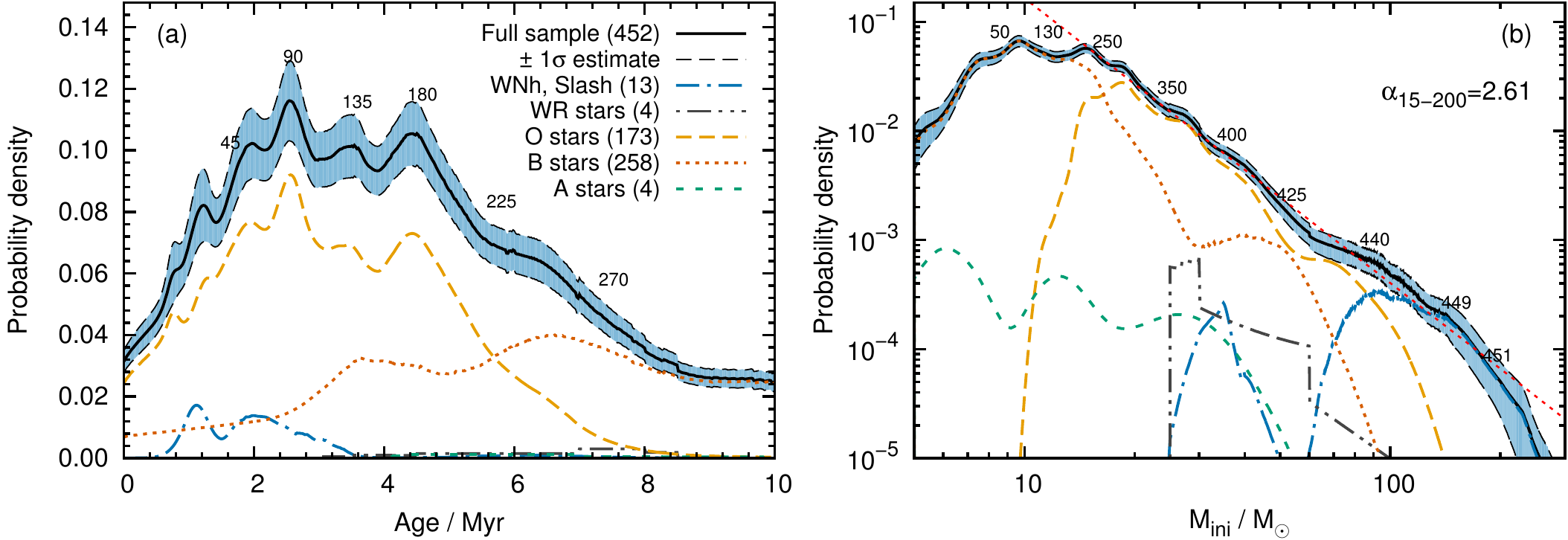}
\par\end{centering}
\caption{As Fig.~\ref{fig:age-mass-distr-whole-region} but without applying the completeness corrections described in Sect.~\ref{sec:inferring-age-mass-distributions}.}
\label{fig:age-mass-distr-whole-region-no-completeness-correction}
\end{figure*}

Looking at the different stellar types separately, there is the tendency that evolutionary masses are slightly larger than spectroscopic masses ($\Delta M/\sigma_{\Delta M}\,{>}\,0$) in O dwarfs ($\Delta M/\sigma_{\Delta M}\,{=}\,0.25\,{\pm}\,0.17$), O (super)giants ($\Delta M/\sigma_{\Delta M}\,{=}\,0.15\,{\pm}\,0.26$) and B (super)giants ($\Delta M/\sigma_{\Delta M}\,{=}\,0.15\,{\pm}\,0.27$), but it is only significant at the 95\% confidence level in the O dwarf sample. In O stars without luminosity class ($\Delta M/\sigma_{\Delta M}\,{=}\,-0.17\,{\pm}\,0.46$) and B dwarfs ($\Delta M/\sigma_{\Delta M}\,{=}\,-0.11\,{\pm}\,0.13$), the tendency is opposite and the spectroscopic masses appear on average larger than evolutionary masses; however, this is not significant at 95\% confidence. Dividing stars into sub-samples according to their spectroscopic classification can introduce biases because there is a strong correlation of the mass of stars with spectral class that can lead to misleading results as discussed by \citet{2017A&A...601A..79S}. It is therefore advisable to study samples of stars as a whole. In our case of VFTS stars, we find no evidence for a statistically significant mass discrepancy in our full sample.

\section{Age and mass distribution without completeness corrections}\label{sec:am-no-completeness-correction}

To demonstrate the influence of the applied completeness corrections (Sect.~\ref{sec:completeness-corrections}), we compute age and mass distributions of our sample stars without them (Fig.~\ref{fig:age-mass-distr-whole-region-no-completeness-correction}). In comparison to the age and mass distributions that take the completeness corrections into account (Fig.~\ref{fig:age-mass-distr-whole-region}), the changes are small, leaving our conclusions untouched.

\section{Age of the \tld star cluster in NGC~2060}\label{sec:tld1}

\begin{table*}
\centering
\caption{\label{tab:tld1}Stellar parameters of eight OB stars from \citet{1999AJ....118.1684W} in \tld. Given are the \citet{1988Msngr..54...43T} identifiers extended by \citet{1999AJ....118.1684W}, spectral types (SpT), effective temperatures ($T_\mathrm{eff}$), luminosities ($\log L/\lsun$), initial masses ($M_\mathrm{ini}$), initial rotational velocities ($v_\mathrm{ini}$), ages and present-day masses ($M_\mathrm{present}$).}
\begin{tabular}{llcccccc}
\toprule 
 Tes ID & SpT & $\log L/\lsun$ & $T_\mathrm{eff}$ & $M_\mathrm{ini}$ & $v_\mathrm{ini}$ & Age & $M_\mathrm{present}$ \\
  & & & (kK) & ($\msun$) & ($\mathrm{km}\,\mathrm{s}^{-1}$) & (Myr) & ($\msun$) \\
\midrule
\midrule 
2A & O4 III(f)p & $5.82\pm0.10$ & $42.4\pm2.0$ & $56.0^{+7.8}_{-6.6}$ & $100^{+112}_{-61}$ & $2.0^{+0.4}_{-0.4}$ & $53.8^{+6.8}_{-6.3}$ \\
3 & O7--O8 II & $5.59\pm0.12$ & $34.4\pm3.0$ & $38.4^{+6.3}_{-5.1}$ & $100^{+117}_{-63}$ & $3.3^{+0.6}_{-0.5}$ & $37.0^{+5.4}_{-4.8}$ \\
7 & O7.5 V((f)) & $5.41\pm0.10$ & $36.9\pm2.0$ & $33.2^{+3.6}_{-3.3}$ & $100^{+121}_{-63}$ & $3.4^{+0.5}_{-0.6}$ & $32.2^{+3.5}_{-2.9}$ \\
1B & O7 V & $5.40\pm0.10$ & $37.9\pm2.0$ & $33.4^{+3.8}_{-3.1}$ & $100^{+121}_{-63}$ & $3.2^{+0.6}_{-0.6}$ & $32.6^{+3.5}_{-2.9}$ \\
2B & O8 III & $5.21\pm0.10$ & $34.0\pm2.0$ & $26.4^{+2.6}_{-2.4}$ & $100^{+130}_{-65}$ & $4.4^{+0.7}_{-0.7}$ & $25.8^{+2.6}_{-2.2}$ \\
5 & O7 V (N strong) & $5.13\pm0.10$ & $37.9\pm2.0$ & $27.4^{+2.6}_{-2.5}$ & $100^{+142}_{-67}$ & $3.0^{+0.9}_{-1.5}$ & $27.0^{+2.7}_{-2.3}$ \\
4 & O9.5 V & $5.04\pm0.10$ & $32.9\pm2.0$ & $22.6^{+2.0}_{-2.0}$ & $100^{+135}_{-65}$ & $5.1^{+0.9}_{-1.0}$ & $22.2^{+2.1}_{-1.8}$ \\
6 & B0.2 V & $4.73\pm0.10$ & $30.3\pm2.0$ & $17.2^{+1.6}_{-1.3}$ & $100^{+142}_{-66}$ & $6.8^{+1.5}_{-1.8}$ & $17.2^{+1.5}_{-1.3}$ \\
\bottomrule
\end{tabular}
\end{table*}

\begin{figure}
\begin{centering}
\includegraphics[width=0.47\textwidth]{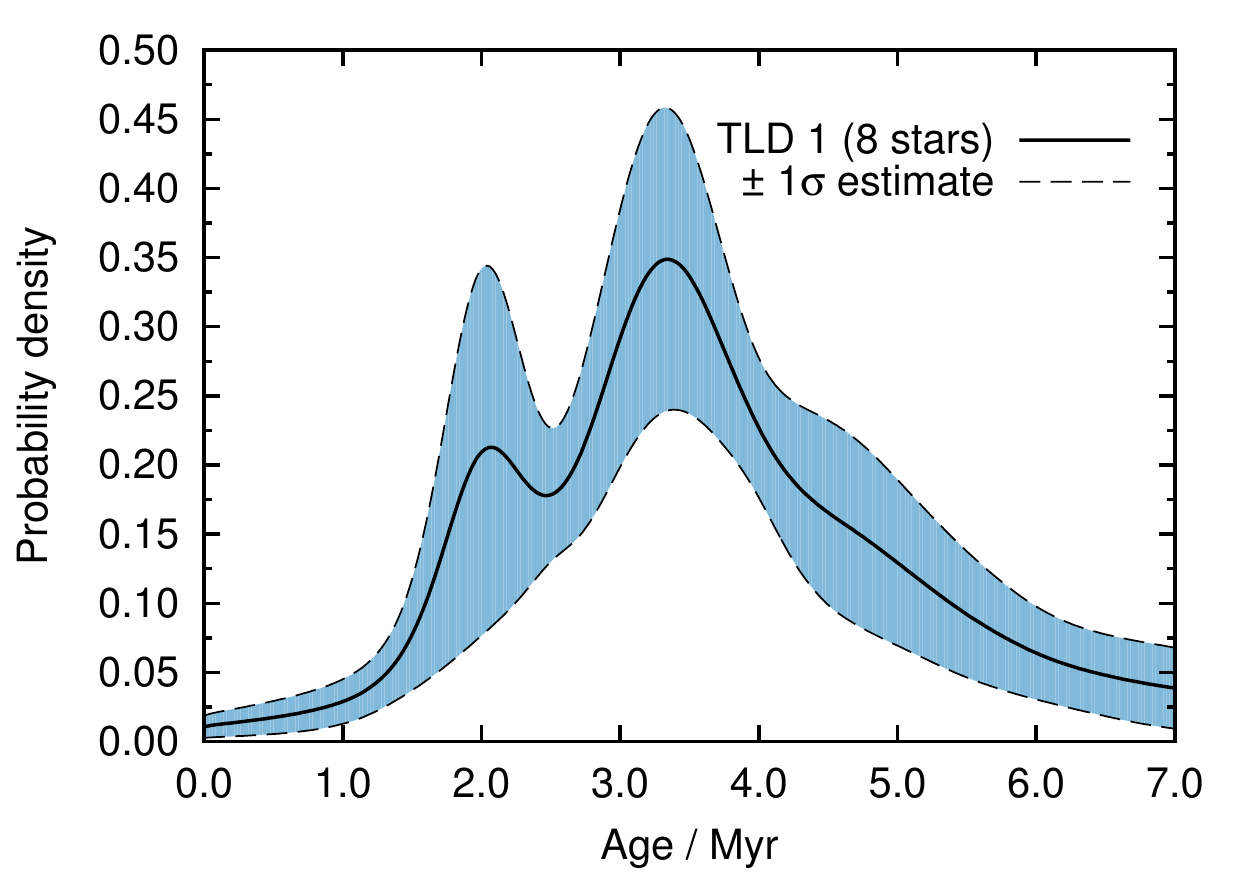}
\par\end{centering}
\caption{Age distribution of eight OB stars in \tld. As before, the blue-shaded region is a bootstrapped $1\sigma$ estimate.}
\label{fig:age-distr-tld1}
\end{figure}

As described in Sect.~\ref{sec:am-ngc2060-region}, we use the spectral classes of \citet{1999AJ....118.1684W} and the calibrations used by \citet{2016MNRAS.458..624C} for massive stars in the R136 cluster core to derive effective temperatures and luminosities of eight OB stars in \tld. Using \bonnsai with the same prior distributions as for the other stars studied here (see Sect.~\ref{sec:stellar-parameters}), we obtain the individual stellar ages and masses, and thereby also an age distribution for stars in the \tld cluster. A summary of the derived stellar parameters of the eight OB stars is provided in Table~\ref{tab:tld1} and the age distribution in Fig.~\ref{fig:age-distr-tld1}. We note that we have no observational constraints on the rotation rates of the stars such that the inferred initial rotational velocities are more or less solely given by the applied prior distribution of the initial rotational velocity.

The age distribution peaks at about $3.3\,\myr$ and suggests that stars in \tld are generally coeval. There is one star, Tes~2A, that is apparently younger than the other stars and two seemingly older stars, Tes~4 and~6. Tes~2A is at the same time also one of the most massive stars in \tld, making it a good candidate for a rejuvenated binary product \citep[either a merger or a product of stable mass transfer; \eg][]{2014ApJ...780..117S,2015ApJ...805...20S}. The apparently old ages of Tes~4 and~6 are not readily understood but we caution that more sophisticated atmosphere modelling is required to obtain more robust stellar parameters (including ages and masses) and hence better constraints on the \tld cluster age and coevality.

\end{document}